\documentclass[A4, English, nofootinbib,superscriptaddress, aps, prd, 10pt]{revtex4-2}
\usepackage[T1]{fontenc}
\usepackage[latin9]{inputenc}
\usepackage{color}
\usepackage{babel}
\usepackage{array}
\usepackage{multirow}
\usepackage{amsmath}
\usepackage{graphicx}
\usepackage{amssymb}
\usepackage{bm}
\usepackage[normalem]{ulem}
\usepackage[unicode=true,pdfusetitle,
 bookmarks=true,bookmarksnumbered=false,bookmarksopen=false,
 breaklinks=false,pdfborder={0 0 0},pdfborderstyle={},backref=false,colorlinks=true]
 {hyperref}
\hypersetup{
 linkcolor=blue, citecolor=cyan}

\makeatletter

\usepackage{amsthm}
\usepackage{xcolor}
\usepackage{soul}

\makeatother

\begin{document}
\title{Is the shear induced spin polarization non-dissipative?}

\author{Jia-Rong Wang}
\email{jpwjr2021@mail.ustc.edu.cn}
\affiliation{Department of Modern Physics, University of Science and Technology
of China, Anhui 230026, China}

\author{Shuo Fang}
\email{fangshuo@mail.ustc.edu.cn}
\affiliation{Department of Modern Physics, University of Science and Technology
of China, Anhui 230026, China}
\affiliation{Technical University of Munich, TUM School of Natural Sciences, Physics Department, James-Franck-Str. 1, 85748 Garching, Germany}

\author{Di-Lun Yang}
\email{dilunyang@gmail.com}
\affiliation{Institute of Physics, Academia Sinica, Taipei, 11529, Taiwan}
\affiliation{Physics Division, National Center for Theoretical Sciences, Taipei, 106319, Taiwan}

\author{Shi Pu}
\email{shipu@ustc.edu.cn}
\affiliation{Department of Modern Physics, University of Science and Technology
of China, Anhui 230026, China}
\affiliation{Southern Center for Nuclear-Science Theory (SCNT), Institute of Modern
Physics, Chinese Academy of Sciences, Huizhou 516000, Guangdong Province,
China}

\begin{abstract}
The shear-induced polarization plays a crucial role in understanding the local polarization of $\Lambda$ and $\overline{\Lambda}$ hyperons. A key puzzle is whether the shear-induced polarization is non-dissipative or not. In this work, we analyzed the shear-induced polarization and the anomalous Hall effects using the entropy flow and H-theorem introduced from quantum (chiral) kinetic theory. While the shear-induced polarization and the anomalous Hall effect do not directly contribute to the entropy production rate, the perturbations associated with the shear tensor lead to an increase in entropy, similar to the role of the shear tensor in classical kinetic theory.
We also examined these effects within the framework of linear response theory using Zubarev's approach. 
The analysis of time-reversal transformation on the spin current in coordinate space suggests that shear-induced polarization violates time-reversal symmetry and, therefore, should vanish. However, a similar analysis of the Wigner function in phase space does not impose any additional constraints on shear-induced polarization, allowing it to persist in phase space as expected. This discussion indicates that time-reversal analysis in coordinate space alone may not be sufficient to determine whether an effect is dissipative. Furthermore, our analysis based on Zubarev's approach suggests that these effects may indeed possess a dissipative nature.
 These findings highlight the limitations of the current theoretical framework in fully characterizing the dissipative properties of these phenomena. 
\end{abstract}

\maketitle

\section{Introduction}


Recently, the global polarization of $\Lambda$ and $\overline{\Lambda}$ hyperons, first proposed in the pioneering works \citep{Liang:2004ph, Betz:2007kg, Gao:2007bc}, has been observed in relativistic heavy-ion collisions ~\citep{ STAR:2017ckg, STAR:2019erd, ALICE:2019aid}. By employing models based on statistical field theory and analyzing the experimental data, it was found that the quark-gluon plasma created in these collisions represents the most vortical fluid observed to date. The global polarization of hyperons is well described by various phenomenological models \citep{Becattini:2007nd,Becattini:2007sr,Becattini:2013fla,Fang:2016vpj,Karpenko:2016jyx,Xie:2017upb,Li:2017slc,Sun:2017xhx,Shi:2017wpk,Xia:2018tes,Shi:2019wzi,Fu:2020oxj,Lei:2021mvp,Ambrus:2020oiw,Vitiuk:2019rfv}.
In most of these models, the global polarization of hyperons is induced by the thermal vorticity of the fluid, under the assumption of global equilibrium in the system.
In addition to global polarization, experiments have also observed the local polarization of $\Lambda$ and $\overline{\Lambda}$ hyperons. However, the experimental data for local polarization is inconsistent with the predictions of the aforementioned models. Understanding local polarization requires the inclusion of additional hydrodynamic effects beyond global equilibrium, such as polarization induced by the shear viscous tensor, fluid acceleration, and gradients of the chemical potential over temperature \citep{Hidaka:2017auj,Liu:2020dxg,Becattini:2021suc,Liu:2021uhn,Fu:2021pok,Becattini:2021iol,Yi:2021ryh,Ryu:2021lnx,Florkowski:2021xvy,Buzzegoli:2022fxu,Becattini:2022zvf,Palermo:2022lvh,Wu:2022mkr,Fu:2022myl,Fu:2022oup,Palermo:2024tza}. By accounting for all these hydrodynamic effects, the local polarization observed in high-energy collisions has been successfully explained \citep{Palermo:2024tza}. Very recently, it has been observed that the local spin polarization in p + A collisions \citep{CMS:2025nqr} cannot be explained by the combined effects of polarization induced by thermal vorticity and the shear viscous tensor \citep{Yi:2024kwu}.
On the other hand, the spin alignment of vector mesons has also been observed in relativistic heavy-ion collisions \citep{STAR:2022fan, ALICE:2023jad}. Notably, the spin alignment of $\phi$ mesons is significantly larger than expected. This large spin alignment of $\phi$ mesons has been investigated using several theoretical models \citep{Sheng:2019kmk, Sheng:2022wsy, Sheng:2022ffb, Muller:2021hpe, Kumar:2022ylt, Kumar:2023ghs, Xia:2020tyd, Li:2022vmb, Wagner:2022gza, Sheng:2024kgg, Yang:2024ejk,Yang:2024qpy}. One successful approach is based on the effective vector meson strong force field \citep{Sheng:2022wsy, Sheng:2022ffb}. This model successfully describes the transverse momentum dependence of the spin alignment of $\phi$ mesons and has also accurately predicted its rapidity dependence \citep{Sheng:2023urn}.
Interestingly, alternative models attribute the spin alignment to effects induced by the shear viscous tensor \citep{Li:2022vmb, Wagner:2022gza}, rather than relying on effective vector meson strong force fields. Such contributions from hydrodynamical field, however,  were shown to be very small by some follow-up works \citep{Dong:2023cng, Liang:2025hxw, Yin:2025gvl, Fang:2025pzy}. For a more comprehensive discussion of these polarization effects, we refer readers to recent reviews \citep{Gao:2020vbh,Hidaka:2022dmn,Becattini:2024uha} and the references therein.

Shear-induced polarization is key to understanding the local polarization of hyperons and may also play an important role in resolving the puzzles associated with the spin alignment of vector mesons. 
Therefore, it is essential to investigate this effect within a fundamental theoretical framework. Broadly speaking, the shear-induced polarization of hyperons corresponds to a spin current (more precisely, the axial-vector current of fermions in its canonical form) generated by the shear viscous tensor in phase space. It is a well-known fact that the shear viscous tensor, as the symmetric part of the energy-momentum tensor in relativistic hydrodynamics, is dissipative, e.g. see  the early studies by Israel-Stewart \cite{Israel:1979wp} or the modern review \citep{Rocha:2023ilf} and the reference therein.  
It is straightforward to show that the shear viscous tensor contributes to an increase in entropy. Consequently, one might expect the spin current induced by the shear viscous tensor to also be dissipative. However, spin currents are often absent from the entropy production rate, which suggests that shear-induced polarization could potentially be non-dissipative.
This leads to an intriguing question that has garnered significant attention: \emph{Is shear-induced polarization dissipative or non-dissipative?}

In this work, we aim to address the above questions. Since the spin current induced by the shear viscous tensor exists in phase space, it is natural to employ quantum kinetic theory to investigate this problem. The main difficulty in this study lies in the lack of a strict definition of entropy density or entropy flow with quantum corrections. Although some progress has been made in Refs. \cite{Chen:2015gta,Yang:2018lew, VanHerstraeten:2021nce, Allori_2019}, 
it remains unclear how to define entropy density with quantum corrections in the framework of kinetic theory. To avoid these ambiguities in the definition of entropy, we adopt the following strategy. First, we utilize the chiral kinetic theory for massless fermions instead of the quantum kinetic theory for massive fermions. 
For massive fermions, linking the axial current associated with the polarization pseudovector to the entropy production rate is challenging. Consequently, applying entropy analysis to shear-induced polarization in such cases becomes difficult. In chiral kinetic theory, axial-vector current is conserved in the absence of electromagnetic fields, and  directly contributes to the entropy production rate. This approach helps us clarify the underlying puzzles.
Second, we restrict our analysis to entropy density up to the order of $\hbar$, as only the leading-log order of the collisional kernel is known so far \citep{Fang:2022ttm, Wang:2022yli, Lin:2022tma,  Fang:2024vds, Lin:2024zik, Wang:2024lis}. Furthermore, the entropy flow at the second order in the gradient expansion (or $\sim \hbar^2$) in relativistic hydrodynamics also suffers from ambiguities \cite{Israel:1979wp}.
Third, we extend the classical kinetic theory to introduce the H-theorem within the quantum kinetic theory framework and verify its consistency with hydrodynamic approaches. Using this formalism, we then analyze the dissipative nature of shear-induced polarization.

The structure of this work is organized as follows. In Sec. \ref{sec:classical_kinetic}, we briefly review the H-theorem and entropy currents in classical kinetic theory. In Sec. \ref{sec:H-theorem-CKT}, we extend these studies to the chiral kinetic theory. Following that, we apply the H-theorem to the chiral kinetic theory using the relaxation-time approach. In Sec. \ref{sec:Application}, we compute the entropy production rate associated with the shear viscous tensor. In Sec.~\ref{sec:Disspative}, we discuss the shear induced polarization in the linear response theory in Zubarev's approaches. 
We summarize this work and discuss the results in Sec. \ref{sec:summary}.
Throughout this work, we adopt the metric $g_{\mu\nu} = \{+,-,-,-\}$ and the projector $\Theta^{\mu\nu} = g^{\mu\nu} - u^\mu u^\nu$, where $u^\mu$ denotes the fluid velocity.  For an arbitrary tensor $A^{\mu\nu}$, we also introduce its traceless part, 
$A^{\left\langle\mu\nu \right\rangle}=\left[\frac{1}{2}\left(\Theta_{\alpha}^{\mu}\Theta_{\beta}^{\nu}+\Theta_{\beta}^{\mu}\Theta_{\alpha}^{\nu}\right)-\frac{1}{3}\Theta^{\mu\nu}\Theta_{\alpha\beta}\right]A^{\alpha\beta}$.






\section{H theorem and entropy current in classical kinetic theory} 
\label{sec:classical_kinetic}

In this section, we will briefly review the H theorem and entropy current in the classical kinetic theory. 
Let us start from the relativistic Boltzmann equation for fermions, 
\begin{equation}
\frac{d}{dt}f_{p}\equiv\frac{p^{\mu}}{E_{p}}\left(\frac{\partial}{\partial x^{\mu}}-F_{\mu\nu}\frac{\partial}{\partial p_{\nu}}\right)f_{p}=C[f_{p}],\label{eq:BE_01}
\end{equation}
where $f_{p}=f_{p}(t,\mathbf{x},\mathbf{p})$ is the distribution
function for on-shell particles with $p^{\mu}=(E_{p},\mathbf{p})$
and $E_{p}=\sqrt{\mathbf{p}^{2}+m^{2}}$, $C[f]$ stands for the collision
term and $F_{\mu\nu}$ denotes the electromagnetic strength tensor.
We further consider the typical 2-2 scatterings for the collisions,
\begin{eqnarray}
C[f_{p}] & = & \frac{1}{2E_{p}}\int d\Gamma_{kk^{\prime}\rightarrow pp^{\prime}}[f_{k}f_{k^{\prime}}(1-f_{p})(1-f_{p^{\prime}})-(1-f_{k})(1-f_{k^{\prime}})f_{p}f_{p^{\prime}}], \label{eq:collision_term_classical}
\end{eqnarray}
where $d\Gamma_{kk^{\prime}\rightarrow pp^{\prime}}=\frac{d^{3}p^{\prime}}{(2\pi)^{3}(2E_{p})}\frac{d^{3}k}{(2\pi)^{3}(2E_{k})}\frac{d^{3}k^{\prime}}{(2\pi)^{3}(2E_{k^{\prime}})}(2\pi)^{4}\delta(k+k^{\prime}-p-p^{\prime})\left|\mathcal{M}_{kk^{\prime}\rightarrow pp^{\prime}}\right|^{2}$
with $\mathcal{M}_{kk^{\prime}\rightarrow pp^{\prime}}$ being the
amplitude for scatterings. The detailed balance condition gives, $\left|\mathcal{M}_{kk^{\prime}\rightarrow pp^{\prime}}\right|^{2}=\left|\mathcal{M}_{pp^{\prime}\rightarrow kk^{\prime}}\right|^{2}$.

Next, we introduce the global and local equilibrium in the kinetic
theory. It is straightforward to prove that the collision term vanishes when $f_p = f^{(0)}_p$, where $f^{(0)}_p$
is the Fermi-Dirac distribution at thermal equilibrium,
\begin{equation}
    f^{(0)}_p=\frac{1}{e^{(u\cdot p-\mu)/T}+1},
    \label{eq:distribution_eq}
\end{equation}
with $u^{\mu},\mu,T$ being fluid velocity, chemical potential and
temperature, We emphasize
that the distribution function which can make $C[f]=0$ does not have
to be the solution of Boltzmann equation (\ref{eq:BE_01}). Therefore,
it is commonly defined the distribution function for the global and
local equilibrium as follows.
\begin{itemize}
    \item Global equilibrium in kinetic theory: the distribution function which
    makes both collision term $C[f]$ and the free screaming part $df/dt$
    vanish. By inserting $f^{(0)}_p$ into Boltzmann equation (\ref{eq:BE_01})
    in the absence of $F^{\mu\nu}$, it gives the Killing condition at
    global equilibrium \citep{DeGroot:1980dk}
    \begin{equation}
    \partial_{\mu}\left(\frac{u_{\nu}}{T}\right)+\partial_{\nu}\left(\frac{u_{\mu}}{T}\right)=0.
    \end{equation}

    \item Local equilibrium in kinetic theory: the distribution function which
makes $C[f]=0$ but the free streaming part $df/dt\neq 0$. While, $\partial_{\mu}\left(\frac{u_{\nu}}{T}\right)+\partial_{\nu}\left(\frac{u_{\mu}}{T}\right)\neq0$. 
\end{itemize}

The conventional Boltzmann's H function is defined as
\begin{equation}
H[f_{p}]=\int\frac{d^{3}p}{(2\pi)^{3}}\mathcal{H}[f_{p}],\label{eq:H_BE}
\end{equation}
where the kernel $\mathcal{H}$ for fermions is given by 
\begin{equation}
\mathcal{H}[f]=-f\ln f-(1-f)\ln(1-f).
\end{equation}
To provide the physical interpretation of $H$, we first introduce
the energy-momentum tensor $T^{\mu\nu}$ and current $j^{\mu}$ by
using Boltzmann equations, 
\begin{eqnarray}
T^{\mu\nu} & = & \int\frac{d^{3}p}{(2\pi)^{3}E_{p}}p^{\mu}p^{\nu}f_{p},\nonumber \\
j^{\mu} & = & \int\frac{d^{3}p}{(2\pi)^{3}E_{p}}p^{\mu}f_{p}.\label{eq:EMT_01}
\end{eqnarray}
The energy density $\varepsilon$, pressure $P$ and number density
$n$ are given by,
\begin{equation}
\varepsilon=u_{\mu}u_{\nu}T^{\mu\nu},\quad P=-\frac{1}{3}\Theta_{\mu\nu}T^{\mu\nu},\quad n=u_{\mu}j^{\mu}.
\end{equation}
Inserting the $f_{p}=f^{(0)}_p$ into definition in Eq. (\ref{eq:H_BE}),
we then notice that the $H$ is the entropy density at the thermal
equilibrium. Analogous to the current, one can define the entropy current as
\begin{equation}
s^{\mu}=\int\frac{d^{3}p}{(2\pi)^{3}}\frac{p^{\mu}}{E_{p}}\mathcal{H}[f_{p}],\label{eq:classical entropy}
\end{equation}
whose time component is the entropy density $\mathcal{H}$. 
Taking the space-time divergence to $s^\mu$ in the kinetic theory, 
we can compute the covariant derivative of $H$,
\begin{eqnarray}
\partial_\mu s^\mu & = & -\int \frac{d^{3}p}{(2\pi)^3}\ln\left(\frac{f_p}{1-f_p}\right)C[f_p]+\int d^{3}p\ln\left(\frac{f_p}{1-f_p}\right)\frac{p^{\mu}}{E_{p}}F_{\mu\nu}\partial_{p}^{\nu}f_p\nonumber \\
 & = & - \int \frac{d^{3}p}{(2\pi)^3}\ln\left(\frac{f_p}{1-f_p}\right)C[f_p]+2 \int d^{4}p \partial_{p}^{\nu} \left[ \delta(p^{2}-m^{2})\theta(p^{0})\ln\left(\frac{f_p}{1-f_p}\right)p^{\mu}F_{\mu\nu}f_p\right]\nonumber \\
 & = & -\int \frac{d^{3}p}{(2\pi)^3}\ln\left(\frac{f_p}{1-f_p}\right)C[f_p],
 \label{eq:partial_s_classical}
\end{eqnarray}
where we have dropped the surface terms in the integration. Due to the symmetry, we find 
\begin{eqnarray}
\partial_\mu s^\mu  =  \frac{1}{4}\int\frac{d^{3}p}{2E_{p}}d\Gamma_{kk^{\prime}\rightarrow pp^{\prime}}f_{k}f_{k^{\prime}}(1-f_{p})(1-f_{p^{\prime}})\left[(1-X)\ln X \right]\geq0,\quad X>0,\label{eq:dH_BE}
\end{eqnarray}
where 
\begin{equation}
X=\frac{f_{p}f_{p^{\prime}}(1-f_{k})(1-f_{k^{\prime}})}{(1-f_{p})(1-f_{p^{\prime}})f_{k}f_{k^{\prime}}}.\label{eq:def_X}
\end{equation}
We notice that when we take the $f_p=f^{(0)}_p$ then it is straightforward to get
$X\rightarrow1$, and 
the $\partial_\mu s^\mu =0$.

On the other hand, we can also obtain the entropy density satisfying the relativistic extended thermodynamic relation,
\begin{equation}
s^{\mu}=\frac{P}{T}u^{\mu}+\frac{1}{T}u_{\nu}T^{\mu\nu}-\frac{\mu}{T}j^{\mu}, \label{eq:entropy_flow_0}
\end{equation}
by employing $f_{p}=f^{(0)}_p$.
Eq.~(\ref{eq:entropy_flow_0}) is the generalization of the conventional thermodynamic relation, 
$s = (P + \varepsilon - \mu n)/T$, which has been widely discussed in many hydrodynamic studies \cite{Israel:1979wp}. In general, it is not guaranteed that Eq.(\ref{eq:entropy_flow_0}) holds for an arbitrary distribution function $f_p$.
To address this, we focus on systems that are very close to thermal equilibrium, i.e.,
\begin{equation}
    f_p = f^{(0)}_p + \delta f_p,
    \label{eq:f_near_eq}
\end{equation}
where $\delta f_p \sim \mathcal{O}(\partial)$. Under these conditions, Eq.~(\ref{eq:entropy_flow_0}) is valid. In relativistic hydrodynamics, this relation holds up to $\mathcal{O}(\partial^2)$, which is consistent with results from kinetic theory \cite{Israel:1979wp}. 
By using Eq.~(\ref{eq:f_near_eq}), we can insert the expression for energy-momentum tensor and current into Eq.~(\ref{eq:partial_s_classical}) and obtain, 
\begin{equation}
    \partial_\mu s^\mu = \left(T^{\mu\nu} - T^{\mu\nu}_{(0)}\right) \partial_\mu \left( \frac{u_\nu}{T}\right) -\left(j^\mu - j^\mu_{(0)}\right)\partial_\mu \left(\frac{\mu}{T}\right), 
    \label{eq:productation_entropy_classical}
\end{equation}
where $T^{\mu\nu}_{(0)}$ and $ j^\mu_{(0)}$ are the energy-momentum tensor and current derived with $f_p = f^{(0)}_p$. The Eq.~(\ref{eq:productation_entropy_classical}) is consistent with the entropy production rate up to $\mathcal{O}(\partial^2)$ in the relativistic hydrodynamics \cite{Israel:1979wp} and statistical field theory \citep{zubarev1979derivation}. We emphasize that $dH/dt=0$ at both global and local equilibrium in the classical kinetic theory, i.e. the entropy is at its maximum value.

At last, we consider the shear viscous tensor as an example to show how to implement the H theorem in the classical kinetic theory.  
If the system is close to the equilibrium, we can consider the perturbation
around the distribution function at the thermal equilibrium as in Eq.~(\ref{eq:f_near_eq}), 
where one usually assume that 
\begin{equation}
\delta f_{p}=f^{(0)}_p(1-f^{(0)}_p)\chi_{p},\label{eq:delta_f}
\end{equation}
for convenience. Since $f_{p}$ must obey the Boltzmann equation (\ref{eq:BE_01}),
the $\chi_{p}$ can be expressed as the moment coupled to dissipative
effects. Here, we consider the shear viscous tensor as an example.
Inserting Eq.~(\ref{eq:delta_f}) into Eqs. (\ref{eq:EMT_01}), we
can obtain the shear viscous tensor in terms of integration of $\delta f_{p}$
over momentum, 
\begin{eqnarray}
\pi^{\mu\nu} & = & T^{\left\langle \mu\nu\right\rangle }=\int\frac{d^{3}p}{(2\pi)^{3}E_{p}}p^{\left\langle \mu\right.}p^{\left.\nu\right\rangle }f^{(0)}_p(1-f^{(0)}_p)\chi_{p}.\nonumber \\
 & \equiv & 2\eta\partial^{\left\langle \mu\right.}u^{\left.\nu\right\rangle },
 \label{eq:shear_tensor_classical}
\end{eqnarray}
where $\eta$ is shear viscosity. By mapping $\frac{d}{dt}f^{(0)}_p=C[\delta f_{p}]$,
we obtain the terms related to the shear viscous tensor,
\begin{equation}
\chi_{p}=A(\mathbf{p})p^{i}p^{j}\partial_{\left\langle i\right.}u_{\left.j\right\rangle }=A(\mathbf{p})p^{i}p^{j}\frac{\pi_{ij}}{2\eta},
\end{equation}
where $A(\mathbf{p})$ needs to be solved through the following constrain
equation, 
\begin{equation}
f^{(0)}_p(1-f^{(0)}_p)p^{\left\langle i\right.}p^{\left.j\right\rangle }=\frac{T}{2E_{p}}\int d\Gamma_{kk^{\prime}\rightarrow pp^{\prime}}f^{(0)}_k f^{(0)}_{k^\prime} f^{(0)}_p f^{(0)}_{p^\prime} \left[B^{ij}(\mathbf{p})+B^{ij}(\mathbf{p}^{\prime})-B^{ij}(\mathbf{k})-B^{ij}(\mathbf{k}^{\prime})\right],
\end{equation}
with $B^{ij}(\mathbf{k})=A(\mathbf{k})k^{\left\langle i\right.}k^{\left.j\right\rangle }$.
 We can compute the correction to entropy flow,
\begin{equation}
s_{(1)}^{\mu}=\frac{\pi_{ij}}{2\eta}\int\frac{d^{3}p}{(2\pi)^{3}}\frac{p^{\mu}}{E_{p}}\ln\left(\frac{1-f^{(0)}_p}{f^{(0)}_p}\right)A(\mathbf{p})f^{(0)}_p(1-f^{(0)}_p)p^{i}p^{j}+\mathcal{O}(\chi^{2}),
\end{equation}
and entropy production rate in this
case,
\begin{eqnarray}
\partial_{\mu}s_{(1)}^{\mu} = \int\frac{d^{3}p}{(2\pi)^{3}}\frac{p^{\mu}}{E_{p}}\delta f_{p}\partial_{\mu}\ln\left(\frac{1-f^{(0)}_p}{f^{(0)}_p}\right)
= \frac{\partial_{\mu}u_{\nu}}{T}T_{\left(1\right)}^{\mu\nu}
=\frac{\pi^{\mu\nu}\pi_{\mu\nu}}{2\eta T} \geq 0, \label{eq:entropy_prod_shear_classical}
\end{eqnarray}
which shows that shear viscous tensor is dissipative.  

Before concluding this section, we summarize the procedure for constructing the entropy flow in classical kinetic theory. To define the entropy flow $s^\mu$, one replaces the distribution function $f_p$ in the current with the $\mathcal{H}$ function. Next, it is necessary to verify the positivity of the entropy production rate, $\partial_\mu s^\mu$, as defined in kinetic theory. Additionally, one must confirm that the entropy flow, defined through the $\mathcal{H}$ function, satisfies the macroscopic hydrodynamic equation (\ref{eq:entropy_flow_0}) up to $\mathcal{O}(\partial)$.
Finally, we note that $\mathcal{H}$, when constructed from an arbitrary distribution function, does not necessarily represent the entropy density satisfying the thermodynamic relations in Eqs.~(\ref{eq:entropy_flow_0}, \ref{eq:productation_entropy_classical}). 
As a result, we will implement the $H$-theorem when the system is close to equilibrium.

\section{H theorem and entropy current in chiral kinetic equation } 
\label{sec:H-theorem-CKT}

In this section, we extend our study to quantum kinetic theory.
As previously mentioned, the axial-vector current for massless fermions directly corresponds to the spin current without any ambiguities. For simplicity, we will focus on the chiral kinetic theory for massless fermions. More details can be found in the recent review \cite{Hidaka:2022dmn}.

\subsection{Review on the chiral kinetic theory}
We introduce the gauge invariant
Wigner function defined as the Wigner transformation of two point functions for right-handed fermions, \citep{Hidaka:2016yjf},
\begin{equation}
\grave{S}^{<}\left(p,X\right)\equiv\int d^{4}Ye^{\frac{ip\cdot Y}{\hbar}}S^{<}\left(x,y\right),
\end{equation}
where
$S_{ab}^{<}\left(x,y\right)=U\left(y,x\right)\left\langle \chi^\dagger_{b}\left(y\right)\chi_{a}\left(x\right)\right\rangle$ with $\chi$ being the Pauli spinor,
and $U(y,x)$ 
is the gauge link to ensure the gauge invariance. Here we define $X=(x+y)/2$ and $Y=x-y$. The perturbative solution up to $\mathcal{O}\left(\hbar \partial\right)$ 
for Wigner function of right-handed fermions is given by \citep{Hidaka:2016yjf,Hidaka:2022dmn},
\begin{equation}
\grave{S}_{R}^{<\mu}=2\pi\bar{\epsilon}_{(n)}\left[\delta\left(p^{2}\right)\left(p^{\mu}+\hbar S_{\left(n\right)}^{\mu\nu}\mathcal{D}_{\nu}\right)+\hbar\epsilon^{\mu\nu\alpha\beta}p_{\nu}F_{\alpha\beta}\frac{\partial\delta\left(p^{2}\right)}{2\delta p^{2}}\right]f_{p}^{\left(n\right)}, 
\label{eq:S_R_01}
\end{equation}
where  
\begin{equation}
\mathcal{D}_{\nu}f_{p}^{\left(n\right)}=\Delta_{\nu}f_{p}^{\left(n\right)}-\mathcal{C}_{\nu},\;\;\Delta_{\mu}=\partial_{\mu}+F_{\nu\mu}\partial_{p}^{\nu},
\end{equation}
with $\mathcal{C}_{\nu}$ denoting the collisional corrections,  $\bar{\epsilon}_{(n)}$ is the sign of $p\cdot n$ with $n^\mu$ being the frame vector. The general expression for $\mathcal{C}_{\nu}$ can be found in Ref.~\citep{Hidaka:2016yjf} and also see Ref.~\citep{Fang:2024vds} for the $U(1)$ gauge interactions under the hard thermal loop approximation. Here, the spin tensor is given by 
\begin{equation}
S_{\left(n\right)}^{\mu\nu}=\frac{\epsilon^{\mu\nu\alpha\beta}p_{\alpha}n_{\beta}}{2\left(p\cdot n\right)}. \label{eq:spin_tensor_n}
\end{equation} 


Different with Eq.~(\ref{eq:distribution_eq}), the global equilibrium distribution function for chiral kinetic theory contains the quantum corrections \citep{Chen:2015gta,Hidaka:2017auj},
\begin{equation}
f_{p}^{eq}=\frac{1}{e^{g}+1},\;\;\; g=\beta p\cdot u-\beta \mu_{R}+\frac{\hbar S_{\left(u\right)}^{\mu\nu}}{2}\partial_{\mu}\left(\beta u_{\nu}\right),\label{eq:df_in_geq}
\end{equation}
where $\beta\equiv 1/T$ and $\mu_R$ denotes the chemical potential for right-handed fermions. In global equilibrium, the distribution function is frame-independent, and we typically set $n^{\mu} = u^{\mu}$. Similarly, in local equilibrium, we also set $n^{\mu} = u^{\mu}$ for simplicity. 

The canonical energy momentum tensor and current are defined as following, 
\begin{eqnarray}
    T^{\mu\nu}&=&\int\frac{d^{4}p}{\left(2\pi\right)^{4}}\left(p^{\mu}\grave{S}^{<\nu}+p^{\nu}\grave{S}^{<\mu}\right)=u^{\mu}u^{\nu}\epsilon-P\Theta^{\mu\nu}+\Pi_{non}^{\mu\nu}+\Pi_{dis}^{\mu\nu}, \nonumber \\
J^{\mu}&=& 2\int\frac{d^{4}p}{\left(2\pi\right)^{4}}\grave{S}^{<\mu} =n u^{\mu}+\nu_{non}^{\mu}+\nu_{dis}^{\mu},
\end{eqnarray}
where the non-dissipative effects from first-order anomalous transport
are related to the chiral transports,
\begin{equation}
\Pi_{non}^{\mu\nu}=\hbar\xi_{\omega}\left(\omega^{\mu}u^{\nu}+\omega^{\nu}u^{\mu}\right)+\hbar\xi_{B}\left(B^{\mu}u^{\nu}+B^{\nu}u^{\mu}\right),\ \nu_{non}^{\mu}=\hbar\sigma_{B}B^{\mu}+\hbar\sigma_{\omega}\omega^{\mu}.
\end{equation}
Here, the transport coefficients $\xi,\xi_B,\sigma,\sigma_B$ can be computed systematically under quantum kinetic theory, e.g. see Refs.~\citep{Gao:2012ix,Hidaka:2017auj} as an example. The $B^\mu$ is the magnetic fields defined as $
    B^\mu= \frac{1}{2}\epsilon^{\mu\nu\alpha\beta}u_\nu F_{\alpha \beta}$.
The $\Pi_{diss}^{\mu\nu}$ and $\nu_{diss}^\mu$ stand for the dissipative terms derived by inserting the perturbations near the thermal equilibrium $\delta f_{p}^{\left(u\right)}=f_{p}^{\left(u\right)}-f_{p}^{eq}$ into the general expression for $T^{\mu\nu}$ and $J^\mu$. By using the distribution function in the thermal equilibrium, one can also obtain the conservation equations with chiral anomaly,
\begin{equation}
\partial_{\mu}J^{\mu}=-\frac{\hbar}{4\pi^{2}}E_{\mu}B^{\mu},\ \partial_{\mu}T^{\mu\nu}=F^{\nu\mu}J_{\mu}.\label{eq:chiral anomaly}
\end{equation}

By using the solution of Wigner function in Eq.~(\ref{eq:S_R_01}), one can further derive the following kinetic equation for $f_p^{(n)}$ \citep{Hidaka:2016yjf},
\begin{eqnarray}
 &  & \delta\left(p^{2}-\hbar\frac{B\cdot p}{p\cdot u}\right)\left[p\cdot\Delta+\hbar\frac{S_{\left(u\right)}^{\mu\nu}E_{\mu}}{p\cdot u}\Delta_{\nu}+\hbar S_{\left(u\right)}^{\mu\nu}\left(\partial_{\mu}F_{\rho\nu}\right)\partial_{p}^{\rho}+\hbar\left(\partial_{\mu}S_{\left(u\right)}^{\mu\nu}\right)\Delta_{\nu}\right]f_{p}^{\left(u\right)}\nonumber \\
 & = & \delta\left(p^{2}-\hbar\frac{B\cdot p}{p\cdot u}\right)\overline{\mathcal{C}}, \label{eq:CKT}
\end{eqnarray}
where $E^\mu=F^{\mu\nu}u_\nu$ and
\begin{eqnarray} \overline{\mathcal{C}}&=&\left(p^{\mu}+\hbar\frac{S_{\left(u\right)}^{\nu\mu}E_{\nu}}{p\cdot u}+\hbar\partial_{\rho}S_{\left(u\right)}^{\rho\mu}\right)\tilde{\mathcal{C}}_{\mu},\label{eq:C_full}
\end{eqnarray}
is the collision term with $\tilde{\mathcal{C}}_{\mu}$ being a vector related to the self-energies of the theory. We refer Refs. \citep{Fang:2022ttm,  Fang:2024vds} for the systematical discussion on collision terms. Eq.~(\ref{eq:CKT}) is considered as the classical Boltzmann equation (\ref{eq:BE_01}) with quantum corrections. From now on, we will use the Eq.~(\ref{eq:CKT}) as our starting point to investigate the entropy flow instead of the Wigner function.

\subsection{Entropy flow in chiral kinetic theory} \label{sec:entropy_CKT}

Following the same strategy for the classical kinetic theory, we introduce the entropy flow in the chiral kinetic theory by  replacing $f_{p}^{\left(u\right)}$ with the Boltzmann's $\mathcal{H}$ function, 
\begin{equation}
s_{R}^{\mu}=2\int\frac{d^{4}p}{\left(2\pi\right)^{3}}\bar{\epsilon}_{(u)}\left[\delta\left(p^{2}\right)\left(p^{\mu}+\hbar S_{\left(u\right)}^{\mu\nu}\mathcal{D}_{\nu}\right)+\hbar\epsilon^{\mu\nu\alpha\beta}p_{\nu}F_{\alpha\beta}\frac{\partial\delta\left(p^{2}\right)}{2\partial p^{2}}\right]\mathcal{H}\left(f_{p}^{\left(u\right)}\right). 
\label{eq:entropy_flow_01}
\end{equation}
We notice that the above equation has already been assumed in the early studies for Wigner function done by some of us \cite{Yang:2018lew}. Here, for simplicity, we consider the entropy flow for the right-handed fermions only. 

Next, let us examine whether the entropy flow defined in Eq.~(\ref{eq:entropy_flow_01}) is consistent with the one defined in relativistic hydrodynamics.
We can compute the entropy flow at local equilibrium. Inserting $f_p^{eq}$ in Eq.~(\ref{eq:df_in_geq}) with $n^\mu=u^\mu$ into Eq.~(\ref{eq:entropy_flow_01}), yields, 
\begin{equation}
s_{R,leq}^{\mu}=\beta \left(u^{\mu}P+T_{leq}^{\mu\nu}u_{\nu}-\mu_{R}J_{leq}^{\mu}+\xi_{B}B^{\mu}+2   \xi_{\omega}\omega^{\mu}\right),
\end{equation}
which agrees with the entropy flow in the chiral anomalous hydrodynamics \cite{Son:2009tf, Pu:2010as}. The terms $\xi_{B}B^{\mu}+2 \xi_{\omega}\omega^{\mu}$ are quantum correction to the classical entropy flow in the presence of electromagnetic fields. Following the Refs.~\citep{Son:2009tf, Pu:2010as, Yang:2018lew}, one can obtain that 
\begin{equation}
    \partial_\mu s^\mu_{R,leq}=0, \label{eq:partial_s_hydro}
\end{equation}
as expected. We emphasize that to obtain the above result, we have assume that the $\mathcal{C}_\nu[f_{p}^{eq}]=0$ in Eq.~(\ref{eq:entropy_flow_01}).

At last, let us check the positivity of the entropy production rate. For simplicity, we will neglect the background electromagnetic field. 
Similar to Eq.~(\ref{eq:delta_f}), we consider a perturbation in the distribution function near the local equilibrium, 
\begin{equation}
f_{p}^{\left(n\right)}=f_{p}^{eq}+\delta f_{p},\ \delta f_{p}=f_{p}^{eq}\left(1-f_{p}^{eq}\right)\chi_{p},\;\; |\chi_{p}|\ll1. 
\label{eq:delta_f2}
\end{equation}
In the absence of electromagnetic fields, the Eq.~(\ref{eq:CKT}) reduces to
\begin{equation}
\left[p\cdot\partial+\hbar\left(\partial_{\mu}S_{\left(n\right)}^{\mu\nu}\right)\partial_{\nu}\right]f_{p}^{\left(n\right)}=\overline{\mathcal{C}}\left[f_{p}^{\left(n\right)}\right].\label{eq:CKT_RTA}
\end{equation}
The collision term $\overline{\mathcal{C}}$ can be computed for a given microscopic theory, e.g. see Refs.~\citep{Fang:2022ttm, Fang:2024vds} for the one given by the quantum electrodynamics in the hard thermal loop approximation. Here, we focus on the general properties of the entropy flow rather than its specifics within a particular theory. To simplify the collision term, we adopt the relaxation-time approximation. In general, the collision term can be written as
\begin{equation}
\overline{\mathcal{C}}\left[f_{p}^{\left(n\right)}\right]=- \frac{1}{\tau_{R}}\left(p\cdot u+\hbar\frac{p\cdot\mathcal{R}}{\left(p\cdot u\right)^{2}}\right)\delta f_{p}, \label{eq:RTA_01}
\end{equation}
where $\tau_{R}$ is the relaxation time. $\mathcal{R}^{\mu}$ is an arbitrary spacelike vector of $\mathcal{O}(\partial^0)$ 
and, therefore, can be parameterized as 
\begin{equation}
    \mathcal{R}^\mu = c_1 u^\mu + c_2 \overline{p}^\mu  + c_3\overline{R}^\mu, \label{eq:R_parameterization}
\end{equation}
where $\overline{p}^\mu = p_\nu \Theta^{\mu\nu}$ and $\overline{R}^\mu = (\Theta^{\mu\nu} - \overline{p}^\mu \overline{p}^\nu/\overline{p}^2)R_\nu$. Here, $c_{i=1,2,3}$ are at $\mathcal{O}(\partial^0)$.

In the absence of the external field, the entropy current in Eq.~(\ref{eq:entropy_flow_01}) reduces to
\begin{equation}
    s^{\mu}=2\int\frac{d^{4}p}{\left(2\pi\right)^{3}}\bar{\epsilon}_{(u)}\delta\left(p^{2}\right)\left\{\left(p^{\mu}+\hbar S_{\left(u\right)}^{\mu\nu}\partial_{\nu}\right)\mathcal{H}[f_{p}^{\left(n\right)}]-\hbar S_{\left(u\right)}^{\mu\nu}\mathcal{C}_{\nu}
    \left[\mathcal{H}[f_p^{(n)}]\right]
    \right\}.\label{eq:entropy_without_e}
\end{equation}
We can further assume that $\mathcal{C}_\nu\left[\mathcal{H}[f_p]\right]$ is linear to $f_p$ and consider that $s^\mu$ up to $\mathcal{O}(\partial^1)$.  Under these two assumptions, we find that the last terms proportional to $\mathcal{C}_\nu$ will not contribute to the entropy flow. More details can be found in Appendix \ref{sec:appendix_entropy}.

Next, let us consider the entropy production rate. Before doing so, we emphasize that the collisional terms are subject to additional constraints imposed by the macroscopic hydrodynamic equations, $\partial_\mu J^\mu=0$ and $\partial_\mu T^{\mu\nu}=0$, in the current approach. In classical kinetic theory, these hydrodynamic conservation equations can be automatically derived by applying the detailed balance conditions in the collisional kernel, which arise from time-reversal symmetry in microscopic theories (see the discussion around Eq.~(\ref{eq:collision_term_classical})). However, in quantum kinetic theory, the time-reversal symmetry in the collisional term, including quantum corrections, may not be as straightforward. On the other hand, in the relaxation time approximation, the hydrodynamic conservation equations become additional constraints, even in classical kinetic theory. Therefore, in this work, we impose the requirement that the collisional term in chiral kinetic theory satisfies the conditions dictated by the hydrodynamic conservation equations,
\begin{eqnarray}
    0&=&\partial_{\mu}J^{\mu}=2\int\frac{d^{4}p}{\left(2\pi\right)^{3}}\bar{\epsilon}_{(u)}\delta\left(p^{2}\right)\bar{\mathcal{C}}\left[f_{p}^{\left(u\right)}\right], \nonumber \\
    0 & = & u_{\nu}\partial_{\mu}T^{\mu\nu}
   = \int\frac{d^{4}p}{\left(2\pi\right)^{3}}\bar{\epsilon}_{(u)}\delta\left(p^{2}\right)\left\{ 2\left(p\cdot u\right)\bar{\mathcal{C}}\left[f_{p}^{\left(u\right)}\right]\right.\nonumber \\
    &  & \left.-\hbar\left[\left(p\cdot u\right)\left(\partial_{\mu}S_{\left(u\right)}^{\mu\alpha}\right)\partial_{\alpha}+p^{\mu}\left(\partial_{\mu}u_{\nu}\right)S_{\left(u\right)}^{\nu\alpha}\partial_{\alpha}\right]f_{p}^{\left(u\right)}\right\} . \label{eq:constrains_hydro}
\end{eqnarray}
By using the above conditions, we can compute the entropy production rate, 
\begin{equation}
    \partial_{\mu}s^{\mu}=\frac{2\hbar}{\tau_{R}}\int\frac{d^{4}p}{\left(2\pi\right)^{3}}\frac{\bar{\epsilon}_{(u)}}{p\cdot u}\delta\left(p^{2}\right)c_{1}f_{p}^{eq}\left(1-f_{p}^{eq}\right)\chi_{p}^{2},
    \label{eq:entropy_production_rate_03}
    \end{equation}
where $c_1$ is introduced in Eq.~(\ref{eq:R_parameterization}). 
To require $\partial_{\mu}s^{\mu}\geq0$ for an arbitrary initial
state, a physical condition served as a sufficient constraint could
be introduced $c_{1}\geq0$. More details for deriving the above equations can be found in Appendix \ref{sec:appendix_entropy}.

\section{Application to spin transports}
\label{sec:Application}

In this sections, we will implement the entropy flow and entropy production rate introduced in the previous section to analyze the shear induced polarization and anomalous Hall effects induced by shear tensor.


For the massless fermions, the spin vector in the phase space corresponds to the axial-vector current in the Wigner function. Since we focus on the right handed fermions, the axial-vector current reduces to the $\grave{S}_{R}^{<\mu}$ in phase space. Inserting the distribution function in the local equilibrium in Eq.~(\ref{eq:df_in_geq}) into Eq.~(\ref{eq:S_R_01}), we can obtain the spin vector induced by various hydrodynamic effects in the absence of electromagnetic fields \cite{Yi:2021ryh, Hidaka:2017auj,Wu:2022mkr}, 
\begin{equation}
    \grave{S}_{R, leq}^{<\mu} = S^\mu_{\textrm{thermal}} + S^\mu_{\textrm{shear}} + 
    S^\mu_{\textrm{accT}} + S^\mu_{\textrm{chemical}}, 
    \label{eq:spin_vector_leq}
\end{equation}
where the subscripts, "thermal, "shear", "accT and "chemical" stand for
the terms related to the thermal vorticity, shear viscous tensor, fluid acceleration,and gradient of $\beta \mu$,respectively. Their expressions are as follows,
\begin{eqnarray}
    S_{\textrm{thermal}}^{\mu} & = & a\frac{1}{2}\epsilon^{\mu\nu\alpha\beta}p_{\nu}\partial_{\alpha}\frac{u_{\beta}}{T},\nonumber \\
    S_{\textrm{shear}}^{\mu} & = & -a\frac{1}{(u\cdot p)T}\epsilon^{\mu\nu\alpha\beta}p_{\alpha}u_{\beta}p^{\sigma}\partial_{\left\langle\sigma\right.}u_{\left.\nu\right\rangle},\nonumber \\
    S_{\textrm{accT}}^{\mu} & = & -a\frac{1}{2T}\epsilon^{\mu\nu\alpha\beta}p_{\nu}u_{\alpha}\left[(u\cdot \partial)u_{\beta}-\frac{1}{T}\partial_{\beta}T\right],\nonumber \\
    S_{\textrm{chemical}}^{\mu} & = & a\frac{1}{(u\cdot p)}\epsilon^{\mu\nu\alpha\beta}p_{\alpha}u_{\beta}\partial_{\nu}\frac{\mu}{T},
    \label{eq:shear_induced_polarization}
\end{eqnarray}
where $a= 4\pi \hbar \overline{\epsilon}_{(u)}\delta(p^2) f_p^{eq}(1-f_p^{eq})$. Inserting Eq.~(\ref{eq:spin_vector_leq}) into the modified Cooper-Frye formulae \citep{Becattini:2013fla,Fang:2016vpj}, one can derive the spin polarization induced by the thermal vorticity, shear tensor, fluid acceleration and the gradient of $\beta \mu$.

The interaction corrections to the spin vector in phase space have been systematically computed in Refs.~\citep{Lin:2022tma, Fang:2023bbw, Lin:2024zik, Fang:2024vds, Lin:2024svh, Wang:2024lis, Fang:2025pzy, Weickgenannt:2022zxs, Weickgenannt:2022qvh, Weickgenannt:2024ibf, Wagner:2024fry, Sapna:2025yss}. Here, we focus on the interaction corrections related to the shear viscous tensor. As pointed out in Ref.~\citep{Fang:2024vds}, as well as in Refs.~\citep{ Weickgenannt:2022qvh,Lin:2024zik}, one type of interaction correction associated with the shear viscous tensor originates from interactions but does not depend on the coupling constant. Similar terms have been widely discussed in condensed matter physics and are related to the anomalous and spin Hall effect \citep{RevModPhys.82.1539,sinova2015spin,dyakonov2017spin}. To avoid complexity, we reproduce this term using the relaxation-time approximation for simplicity.

Inserting the perturbation of distribution function around the local equilibrium in Eq.~(\ref{eq:delta_f}) into the chiral kinetic equation (\ref{eq:CKT_RTA}) with relaxation-time approaches, we can derive the expression for the $\delta f_p$, 
\begin{equation}
    \delta f_p = \frac{\tau_R}{u\cdot p} f_p^{(0)} (1- f_p^{(0)})
    \beta p^\mu p^\nu \partial_{\left\langle\mu\right.} u_{\left.\nu\right\rangle}+.... , \label{eq:detla_f_RTA}
\end{equation}
where $...$ stands for other  terms irrelevant to shear tensor $\partial_{\left\langle\mu\right.} u_{\left.\nu\right\rangle}$. Inserting $\delta f_p$ into Eq.~(\ref{eq:S_R_01}), we obtain the interaction corrections to 
$\grave{S}_{R}^{<\mu}$,
\begin{eqnarray}
    \delta S^{<,\mu}_R & = & 2\pi p^{\mu}\delta\left(p^{2}\right)\delta f_{p}+2\pi\hbar\delta\left(p^{2}\right)S_{\left(u\right)}^{\mu\alpha}\left(\partial_{\alpha}\delta f_{p}-\mathcal{C}_{\alpha}\left[\delta f_{p}\right]\right).
\end{eqnarray}
Analogous to the discussion in Appendix \ref{sec:appendix_entropy}, we parameterize the $\mathcal{C}_\nu [\delta f_p]$ as
\begin{equation}
    \mathcal{C}_\nu [\delta f_p] =\frac{1}{\tau_R^\prime} (d_1 u_\nu +d_2 p_\nu +d_3 \partial_\nu^p) \delta f_p,
\end{equation}
where $d_{1,2,3}$ and the other relaxation time $\tau_R^\prime$ are at $\mathcal{O}(\partial^0)$. Under the above parametrization, we obtain,
\begin{eqnarray}
    \delta S^{<,\mu}_R &=& 2\pi p^{\mu}\delta\left(p^{2}\right)\frac{\tau_R}{u\cdot p} f_p^{(0)} (1- f_p^{(0)})
    \beta p^\mu p^\nu \partial_{\left\langle\mu\right.} u_{\left.\nu\right\rangle}\nonumber \\ 
    & & -4\pi\hbar \frac{1}{(u\cdot p)}\frac{\tau_R}{\tau_R^\prime}\delta\left(p^{2}\right)S_{\left(u\right)}^{\mu\nu}f_p^{(0)} (1- f_p^{(0)}) c_3 \beta p^\sigma 
    \partial_{\left\langle\nu\right.} u_{\left.\sigma\right\rangle}.
\end{eqnarray} 

In general, $\tau_R$ and $\tau_R^\prime$ can be different. However, in certain special cases, $\tau_R$ and $\tau_R^\prime$ exhibit the same dependence on the coupling constant. As a result, the term proportional to $\tau_R / \tau_R^\prime$ does not explicitly depend on the coupling constant. For example, see the studies of quantum kinetic theory within the Chapman-Enskog expansion \cite{Fang:2024vds} and references therein. Such a term can be interpreted as the anomalous Hall effect by simply replacing the shear tensor with the electric force (see also related discussions in condensed matter physics \citep{RevModPhys.82.1539,sinova2015spin,dyakonov2017spin}).

Now, we are ready to compute the entropy production rate related to the shear induced polarization and anomalous Hall effects. By using the $\delta f_p$ in Eq.~(\ref{eq:detla_f_RTA}), the entropy flow related to the shear tensor, namely $\delta s^\mu$, can be cast into a simple form, 
\begin{equation}
    \delta s^{\mu}=\beta u_{\nu}\delta T^{\mu\nu}-\beta\mu_{R}\delta J^{\mu},\label{eq:delta_entropy_CKT}
\end{equation} 
where
\begin{eqnarray}
    \delta T^{\mu\nu} & = & \int\frac{d^{4}p}{\left(2\pi\right)^{4}}\left(p^{\nu}\delta S_{R}^{<,\mu}+p^{\mu}\delta S_{R}^{<,\nu}\right)\nonumber \\
 & = & \frac{4}{15}\tau_{R}\beta\partial^{\left\langle\mu\right.} u^{\left.\nu\right\rangle}\int\frac{d^{4}p}{\left(2\pi\right)^{3}}\delta\left(p^{2}\right)\left(p\cdot u\right)^{3}f_{p}^{\left(0\right)}\left(1-f_{p}^{\left(0\right)}\right), \label{eq:delta_T}
\end{eqnarray}
and 
\begin{eqnarray}
    \delta J^\mu = 2\int\frac{d^{4}p}{\left(2\pi\right)^{4}}\delta S_{R}^{<,\mu}
 = 2\tau_{R}\beta\partial_{\left\langle\rho\right.}u_{\left.\sigma\right\rangle}\int\frac{d^{4}p}{\left(2\pi\right)^{3}\left(p\cdot u\right)}\delta\left(p^{2}\right)f_{p}^{\left(0\right)}\left(1-f_{p}^{\left(0\right)}\right)p^{\mu}p^{\rho}p^{\sigma}=0. \label{eq:delta_J}
\end{eqnarray}
We can further compute the entropy production rate,
\begin{eqnarray}
    \partial_\mu s^\mu &=&  \delta T^{\mu\nu} \partial_\mu (\beta u_\nu) - \delta J^\mu \partial_\mu (\beta \mu)\nonumber \\
    &=&
    \frac{4}{15}\tau_{R}\beta^{2}\partial_{\left\langle\mu\right.} u_{\left.\nu\right\rangle} \partial^{\left\langle\mu\right.} u^{\left.\nu\right\rangle} \int\frac{d^{4}p}{\left(2\pi\right)^{3}}\delta\left(p^{2}\right)\left(p\cdot u\right)^{3}f_{p}^{\left(0\right)}\left(1-f_{p}^{\left(0\right)}\right)\geq0. \label{eq:productation_entropy_CKT}
\end{eqnarray}
The Eq.~(\ref{eq:productation_entropy_CKT}) is a natural extension of the entropy production rate up to $\mathcal{O}(\partial^2)$ in Eq.~(\ref{eq:productation_entropy_classical}).

Now, let us analyze the contributions of $\delta T^{\mu\nu}$ and $\delta J^\mu$ to $\partial_\mu s^\mu$. Comparing with Eq.(\ref{eq:shear_tensor_classical}), we observe that $\delta T^{\mu\nu}$ corresponds to the shear viscous tensor in classical kinetic theory within the relaxation-time approximation. Accordingly, the entropy production rate $\partial_\mu s^\mu$ is nothing new and corresponds to Eq.(\ref{eq:entropy_prod_shear_classical}) in the same approximation. However, we also note that the terms related to shear induced polarization and the anomalous Hall effect vanish in $\delta J^\mu$, as well as in $\partial_\mu s^\mu$ in the end, after momentum integration.

At this point, we encounter a dilemma. On the one hand, the entropy production rate is nonzero when $\partial_{\left\langle\mu\right.} u_{\left.\nu\right\rangle} \partial^{\left\langle\mu\right.} u^{\left.\nu\right\rangle} \neq 0$, which implies that the effects related to the shear tensor are dissipative. On the other hand, all spin transport terms related to the shear tensor do not contribute to the entropy production rate, suggesting that they are non-dissipative at least for momentum-integrated spin current.

\section{Dissipative nature of local equilibrium Wigner function}
\label{sec:Disspative}

In the previous section, we show that entropy analysis within the framework of quantum kinetic theory might be inadequate for studying shear-induced polarization and the anomalous Hall effect, as these phenomena do not contribute to entropy flow after momentum integration. Consequently, it is essential to explore alternative approaches for analyzing entropy directly in phase space, rather than limiting the analysis exclusively to coordinate space.
In this section, we analyze the shear-induced polarization using linear response theory within Zubarev's approach \citep{zubarev1973nonequilibrium,zubarev1979derivation,van1982maximum}. In Sec.~\ref{subsec:Time-reversal-symmetry-analysis}, we perform a time-reversal symmetry analysis for the leading-order axial-vector Wigner function from linear response theory based on Zubarev's density operator. 
In Sec.~\ref{subsec:Derivation-of-shear-induced}, we explicitly compute the axial-vector Wigner function using such linear response theory, incorporating hydrodynamic-gradient sources up to the first-order gradient. Our results suggest that the axial-vector Wigner function, induced by shear-stress tensor and gradient of $\beta \mu$, could be dissipative when the global equilibrium density operator is chosen as the basis density operator. Different from our previous discussions, in this section we focus
on the general massive fermion fields. 


\subsection{Time-reversal symmetry analysis}
\label{subsec:Time-reversal-symmetry-analysis}

The time reversal transformation $\mathcal{T}$ is an anti-unitary
operator which reverses the temporal component, 
\begin{eqnarray}
\mathcal{T}:(t,\boldsymbol{x}) & \to & (-t,\boldsymbol{x}).
\end{eqnarray}
Under $\mathcal{T}$ transformation, both the momentum and spin are flipped.
It preserves the phase factor $\mathcal{T}e^{ip\cdot x}=e^{ip\cdot x}$
and chemical potential. 
We notice that under the $\mathcal{T}$ transformation, gradient
of chemical potential $\partial_\nu(\beta \mu)$ transforms as 
$\mathcal{T}\partial_{\nu}\alpha  =  -(-1)^{\nu}\partial_{\nu}\alpha$.

For ordinary transport phenomena, the dissipative nature of a transport effect can often be known by using time-reversal symmetry analysis of its transport coefficient. If the transport coefficient is $\mathcal{T}$-odd, the effect is dissipative, whereas if it is $\mathcal{T}$-even, the effect is non-dissipative. For instance, consider the charge current $\boldsymbol{j}^i = \sigma \boldsymbol{E}^i$. The ordinary electrical conductivity $\sigma$ is $\mathcal{T}$-odd, meaning that electrical conduction is a dissipative process. This simple analysis based on time-reversal symmetry is directly linked to the entropy analysis in coordinate space mentioned earlier.
In the above example, the contribution from the current to the entropy production rate is $\partial_\mu s^\mu \sim \frac{1}{T} \boldsymbol{j} \cdot \boldsymbol{E} = \frac{1}{T} \sigma \boldsymbol{E}^2$, which is analogous to Eq.~(\ref{eq:productation_entropy_CKT}), replacing $\partial_\mu(\beta \mu)$ with $E_\mu / T$. Since $\partial_\mu s^\mu$ is $\mathcal{T}$-odd and $\sigma$ is also $\mathcal{T}$-odd, the entropy principle remains valid under time-reversal transformation.
However, this simple analysis does not always hold. Consider a Hall current $\boldsymbol{j}^i = \sigma^{ij} E^j$ with $\sigma^{ij} = -\sigma^{ji}$. Here, the coefficient $\sigma^{ij}$ is also $\mathcal{T}$-odd, but this type of current does not contribute to entropy production. Specifically, $\partial_\mu s^\mu \sim \frac{1}{T} \boldsymbol{j} \cdot \boldsymbol{E} = \frac{1}{T} \sigma^{ij} \boldsymbol{E}^i \boldsymbol{E}^j = 0$. In condensed matter physics, such Hall currents are considered non-dissipative \citep{RevModPhys.82.1539}.

Next, let us apply time-reversal symmetry to analyze the spin current operator $\hat{J}_5^\mu$. 
It appears that time-reversal symmetry would imply the vanishing of shear-induced polarization. However, Eq.~(\ref{eq:shear_induced_polarization}) demonstrates that shear-induced polarization can persist in phase space and plays a crucial role in understanding the spin polarization of hyperons. Therefore, it becomes necessary to examine time-reversal symmetry directly in phase space to determine whether it imposes similar constraints as those derived for $J_5^\mu$.

We start from the linear response theory in the Zubarev's approaches \citep{zubarev1979derivation,Becattini:2019dxo}. 
The global equilibrium density operator, which is independent of hypersurface and chosen as the reference state, is defined as,
\begin{eqnarray}
\hat{\rho}_{\text{GE}} & = & \frac{1}{Z}\exp\left[-\beta_{\nu}\hat{P}^{\nu}+\alpha\hat{Q}+\frac{1}{2}\hat{\mathcal{J}}^{\mu\nu}\varpi_{\mu\nu}\right],
\label{eq:Geq_DO}
\end{eqnarray}
where $\{\hat{P}^{\mu},\hat{Q},\hat{\mathcal{J}}^{\mu\nu}\}$ are
the momentum, $U(1)$ charge and total angular momentum operators, respectively
and $\{\beta_{\mu},\alpha,\varpi_{\mu\nu}\}$ are the according Lagrangian
multipliers. Here one can define the thermodynamical quantities: temperature
$T=1/\sqrt{\beta^{\mu}\beta_\mu}$, chemical potential $\mu=\alpha T$ and spin
chemical potential $\omega_{\mu\nu}=T\varpi_{\mu\nu}$. And the physical
quantities can be defined by their densities after integration over
its surrounding region, 
\begin{eqnarray}
\hat{P}^{\mu}=\int_{\Sigma_{\tau}}{\rm d}\Sigma_{\nu}(x)\hat{T}^{\nu\mu}(x),\; & \hat{Q}=\int_{\Sigma_{\tau}}{\rm d}\Sigma_{\mu}(x)\hat{j}^{\mu}(x),\; & \hat{\mathcal{J}}^{\mu\nu}=\int_{\Sigma_{\tau}}{\rm d}\Sigma_{\alpha}(x)\hat{J}^{\alpha\mu\nu}(x),
\end{eqnarray}
with $\{\hat{T}^{\nu\mu}(x),\hat{j}^{\mu}(x),\hat{J}^{\alpha\mu\nu}(x)\}$
their corresponding density operators. The partition function guarantees
the normalization of $\hat{\rho}_{{\rm GE}}$ such that ${\rm Tr}\hat{\rho}_{{\rm GE}}=1$.
In general the whole non-equilibrium density operator $\hat{\rho}$
can be defined in a one-parameter family spacelike hypersurface $\Sigma_{\tau}$
which is parameterized by $\tau$ defined as the proper time along
the integral curves of the normal vector $n_{\mu}(x)$ \citep{van1982maximum}.
The multipliers $\{\beta_{\mu},\alpha,\varpi_{\mu\nu}\}$ can
be promoted to be spacetime dependent. In such case, the local equilibrium
modification to Eq.~(\ref{eq:Geq_DO}) induced by fluid field gradients
can be obtained by generalizing the quantum Liouville equation:
\begin{eqnarray}
\delta\hat{\rho}_{\text{GE}}(\tau) & = & \lim_{\varepsilon\to0}\int_{-\infty}^{\tau}{\rm d}\tau^{\prime}e^{\varepsilon(\tau^{\prime}-\tau)}\int_{\Sigma_{\tau^{\prime}}}{\rm d}\Sigma_{\lambda}(x)n^{\lambda}\left[\int_{0}^{1}{\rm d}z\hat{\rho}_{{\rm GE}}^{z}\left(\hat{T}^{\nu\mu}(x)\partial_{\nu}\beta_{\mu}(x)-\hat{j}^{\nu}(x)\partial_{\nu}\alpha(x)\right)\hat{\rho}_{{\rm GE}}^{1-z}\right.\nonumber \\
 &  & \left.-\left(\langle\hat{T}^{\nu\mu}(x)\rangle_{{\rm GE}}\partial_{\nu}\beta_{\mu}(x)-\langle\hat{j}^{\nu}(x)\rangle_{{\rm GE}}\partial_{\nu}\alpha(x)\right)\hat{\rho}_{{\rm GE}}\right].\label{eq:Off-GE_DO}
\end{eqnarray}
A physical quantity is defined as $O=\langle\hat{O}\rangle={\rm Tr}(\hat{\rho}\hat{O})$
with $\hat{\rho}=\hat{\rho}_{\text{GE}}+\delta\hat{\rho}_{\text{GE}}$
and $\langle\hat{O}\rangle_{{\rm GE}}={\rm Tr}(\hat{\rho}_{{\rm GE}}\hat{O})$.
Eq.(\ref{eq:Off-GE_DO})
can be further simplified if choosing the hypersurface as ${\rm d}\Sigma_{\mu}=({\rm d}^{3}\boldsymbol{x},0)$
with $n^{\mu}=(1,\boldsymbol{0})$ \citep{Becattini:2019dxo},
\begin{eqnarray}
 &  & O(x)-\langle\hat{O}(x)\rangle_{{\rm GE}}\nonumber \\
 & = & \lim_{k\to0}{\rm Im}\frac{i}{\beta(x)k_{0}}\int_{t_{0}}^{t}{\rm d}^{4}x^{\prime}e^{-ik\cdot(x^{\prime}-x)}\left\langle -[\hat{O}(x),\hat{T}^{\mu\nu}(x^{\prime})]\partial_{\mu}\beta_{\nu}(x)+[\hat{O}(x),\hat{j}^{\mu}(x^{\prime})]\partial_{\mu}\alpha(x)\right\rangle _{{\rm GE}}+\mathcal{O}(\partial^{2}).\nonumber \\
\label{eq:LRT_Leq_corrections}
\end{eqnarray}
It is worth noting that the corrections from Eq.~(\ref{eq:Off-GE_DO}) or (\ref{eq:LRT_Leq_corrections}) are deviations from their local equilibrium quantities. Generally, such a local-equilibrium baseline can be evaluated using a hypersurface-dependent local-equilibrium density operator. For instance, a shear-induced axial-vector Wigner function with explicit $\hat{t}^\mu$ dependence was evaluated in  Ref.~\citep{Becattini:2021suc} and recently reproduced by \citep{Li:2025pef}. In this section, we adopt the simplest case by using the global equilibrium as the baseline. In this case, the polarization arising from the shear-stress tensor is purely from the dissipative source and remains independent of an extra time-like vector $\hat{t}^\mu$.

We emphasize that the correction coming from Eq.~(\ref{eq:Off-GE_DO}) or (\ref{eq:LRT_Leq_corrections}) induces positive entropy production
and is thus dissipative \citep{zubarev1973nonequilibrium}. 
Eq.(\ref{eq:LRT_Leq_corrections}) was originally derived using Zubarev's density operator, also see the discussion in the early pioneering work \citep{zubarev1973nonequilibrium} and later revisited in Refs. \citep{zubarev1979derivation,van1982maximum,Hosoya:1983xm}. A re-derivation of this result, similar to our Eq.~(\ref{eq:Off-GE_DO}), can also be found in Eqs.~(30) and (31) of a recent review \citep{Becattini:2019dxo}.


The representation of the time reversal operator $\mathcal{T}$
acts on a field $\Psi(x)$ as
\begin{eqnarray}
\mathcal{T}\Psi(x)=\hat{T}^{-1}\Psi(x)\hat{T} & = & D(\mathcal{T})\Psi(\mathcal{T}x),
\end{eqnarray}
where $D(\mathcal{T})$ is the representation of $\mathcal{T}$ for a given field operator.
For the fermion axial-vector
operator, we find that $ \hat{T}^{-1}\left[\overline{\psi}(x)\gamma^{5}\gamma_{\mu}\psi(x)\right]\hat{T}  =  (-1)^{\mu}\overline{\psi}(-x_0,\boldsymbol{x})\gamma^{5}\gamma_{\mu}\psi(-x_0,\boldsymbol{x})$, i.e.
\begin{eqnarray}
    \hat{T}^{-1}\hat{J}_{5}^{\mu}(x)\hat{T} = (-1)^{\mu}\hat{J}_{5}^{\mu}(-x_{0},\boldsymbol{x}).
    \label{eq:T-transformation-J5}
\end{eqnarray}
where the temporal part is $\mathcal{T}$-odd while the spatial part is $\mathcal{T}$-even. In global equilibrium, 
$J_{5}^{\mu}(x)=\left\langle\overline{\psi}(x)\gamma^{5}\gamma^{\mu}\psi(x)\right\rangle$
does not receive any corrections from the hydrodynamic gradients such as shear viscous tensor and $\partial_\mu\alpha$, except for
(thermal) vorticity. When the system is slightly driven away from the global-equilibrium
state, by using Eq.~(\ref{eq:LRT_Leq_corrections}), the modification of $J_{5}^{\mu}$ from, e.g. $\partial_\mu \alpha$, reads 
\begin{eqnarray}
\delta J_{5,\alpha}^{\mu}(x) & = & \lim_{k\to0}{\rm Im}\frac{i}{\beta(x)k_{0}}\int_{-\infty}^{t}{\rm d}^{4}x^{\prime}e^{-ik\cdot(x^{\prime}-x)}\left\langle [\hat{J}_{5}^{\mu}(x),\hat{j}^{\lambda}(x^{\prime})]\right\rangle _{{\rm GE}}\partial_{\lambda}\alpha(x).
\end{eqnarray}
Under time-reversal transformation, $\delta J_{5,\alpha}^{\mu}(x) $ follows 
\begin{eqnarray}
\hat{T}^{-1}\delta J_{5,\alpha}^{\mu}(x)\hat{T} & = & -(-1)^{\mu}\delta J_{5,\alpha}^{\mu}(-x_{0},\boldsymbol{x}). \label{eq:Time_reversal_delta_J5}
\end{eqnarray}
Comparing the above equation with Eq.~(\ref{eq:T-transformation-J5}), it implies that
\begin{eqnarray}
\delta J_{5,\alpha}^{\mu}(x) & = & 0, \label{eq:T-symmetry-alpha-source}
\end{eqnarray}
which means the modification of $J_{5}^{\mu}$ from $\partial_\mu \alpha$
is forbidden by time-reversal symmetry. 
Therefore, $J_{5}^{\mu}(x)$ seems to be non-dissipative. 
A similar analysis also holds for the shear-stress source. However, such a chemical-potential gradient originates from the deviation of an equilibrium state, which also suggests that the correction may be dissipative. It indicates that the $\mathcal{T}$-symmetry analysis might not be sufficient to determine whether the transport phenomena in phase space are dissipative or not. Let us turn to the phase-space current for an explicit illustration.

The axial-vector Wigner function is defined as the Wigner transformation of the two point functions, 
\begin{eqnarray}
\mathcal{A}^{<,\mu}(q,X) & = & -\frac{1}{4}\int{\rm d}^{4}Ye^{iq\cdot Y}{\rm Tr}\left(\hat{\rho}\overline{\psi}(y)\gamma^{5}\gamma_{\mu}\psi(x)\right),
\end{eqnarray}
where $\hat{\rho}$ denotes the density operator. Since the density operator is $\mathcal{T}$-even, then $\mathcal{A}^{<,\mu}(q,X)$
 transforms as 
\begin{equation}
    \hat{T}^{-1}\mathcal{A}^{<,\mu}(q,X)\hat{T}
   = -(-1)^{\mu}\mathcal{A}^{<,\mu}(-q_{0},\boldsymbol{q};-X_{0},\boldsymbol{X}),\label{eq:Time_reversal_Axial_WF}
\end{equation}
it is also straightforward to recover Eq.~(\ref{eq:T-transformation-J5}) by using $J_{5}^{\mu}(X) = 4\int\frac{{\rm d}^{4}q}{(2\pi)^{4}}\mathcal{A}^{<,\mu}(q,X)$.
More details for Eq.~(\ref{eq:Time_reversal_Axial_WF}) and for $J^\mu_5$ can be found in Eqs.~(\ref{eq:Time_reversal_Axial_WF_detials}, \ref{eq:Time_reversal_J5_details}) of Appendix \ref{sec:A_under_T}.

The $\mathcal{A}^{\mu}$ with global equilibrium density operator gives the well-known polarization induced by thermal vorticity \citep{Buzzegoli:2017cqy} and its radiative corrections \citep{Fang2:2025}. Its modification
from gradient of chemical potential reads
\begin{eqnarray}
\delta\mathcal{A}_{{\rm LE},\alpha}^{<,\mu}(q,X) & = & \partial_{\nu}\alpha(X)\lim_{k\to0}{\rm Im}\frac{1}{\beta(X)k_{0}}\frac{1}{4}\int{\rm d}^{4}Ye^{iq\cdot Y}\int_{-\infty}^{+\infty}{\rm d}^{4}x^{\prime}e^{-ik\cdot(x^{\prime}-X)}(-i)\theta(X_{0}-x_{0}^{\prime})\nonumber \\
 &  & \;\times\left\langle \hat{\overline{\psi}}_{b}(y)(\gamma^{5}\gamma^{\mu})_{ba}\hat{\psi}_{a}(x)\hat{j}^{\nu}(x^{\prime})-\hat{j}^{\nu}(x^{\prime})\hat{\overline{\psi}}_{b}(y)(\gamma^{5}\gamma^{\mu})_{ba}\hat{\psi}_{a}(x)\right\rangle _{{\rm GE}},\label{eq:CPG_induced_polar}
\end{eqnarray}
where the subscript $\alpha$ denotes the axial-vector current induced
by gradient of chemical potential. After some calculation shown in Appendix \ref{sec:A_under_T}, we derive that
\begin{eqnarray}
\hat{T}^{-1}\delta\mathcal{A}_{{\rm LE},\alpha}^{<,\mu}(q,X)\hat{T} & = & -(-1)^{\mu}\delta\mathcal{A}_{{\rm LE},\alpha}^{<,\mu}(-q_{0},\boldsymbol{q};-X_{0},\boldsymbol{X}).
\label{eq:delta_A_under_T}
\end{eqnarray}

We emphasize that it is highly non-trivial that Eq.~(\ref{eq:delta_A_under_T}) takes the same form as Eq.~(\ref{eq:Time_reversal_Axial_WF}). A rough estimation of $\hat{J}_5^\mu$ and $\delta J_{5,\alpha}^\mu$ in Eqs.~(\ref{eq:Time_reversal_delta_J5}) and (\ref{eq:T-symmetry-alpha-source}) suggests 
that shear-induced polarization breaks time-reversal symmetry and needs to vanish. However, our results demonstrate that the time-reversal symmetry of the
Wigner function in phase space, i.e. $\mathcal{A}^{<,\mu}$ and $\delta\mathcal{A}_{{\rm LE},\alpha}^{<,\mu}$, 
impose no constraints on the transport coefficients of the phase-space current.
When integrating out the momentum dependence, nevertheless, the off-equilibrium corrections to the current $J_{5}^{\mu}(x)$ from Eq.~(\ref{eq:LRT_Leq_corrections}) should vanish in accordance with $\mathcal{T}$-symmetry analysis, indicating that it is non-dissipative.

\subsection{Shear-induced polarization from linear response theory}
\label{subsec:Derivation-of-shear-induced}

We now explicitly evaluate the dissipative corrections to the phase-space axial-vector current based on Eq.~(\ref{eq:LRT_Leq_corrections}). By calculating the various retarded correlators between $\hat{\mathcal{A}}^{<,\mu}$ and $\{\hat{J}_{\rho\sigma\lambda},\hat{T}_{\alpha\beta},\hat{j}_{\nu}\}$, we demonstrate that it is possible to recover the local equilibrium corrections within the framework of linear response theory \citep{Liu:2021uhn}. This implies the dissipative nature of the local equilibrium axial-vector current, particularly the contributions arising from the shear-stress tensor and chemical potential gradients. On the other hand, we emphasize that our results differ from the shear contributions discussed in Ref.~\citep{Becattini:2021suc}, where the polarization pseudovector was computed through an expansion on a general hypersurface.
Here, we focus exclusively on the $\mathcal{O}(\partial^{1})$ contribution, where $\varpi_{\mu\nu}$ does not play a role in Eq.~(\ref{eq:LRT_Leq_corrections}). For second-order contributions based on a gradient expansion within kinetic theory, see Ref.~\citep{Fang:2024vds} from quantum kinetic theory and Refs.~\cite{Sheng:2024pbw, Zhang:2024mhs, Yang:2024fkn} from Zubarev's approaches.


We focus on the free theory and substituting $\{\hat{j}^{\mu},\hat{T}^{\mu\nu}\}$
of free fermion fields. After some calculation shown in Appendix \ref{sec:A_under_T}, we derive that 
\begin{eqnarray}
    \delta\mathcal{A}_{{\rm LE},\alpha}^{<,\mu}(q,X) & = & 2\pi\delta(q^{2}-m^{2})T(\partial_{\nu}\alpha)\epsilon^{\mu\alpha\nu\beta}q_{\beta}u_{\alpha}\frac{1}{2|q_{0}|}\left\{ -\partial_{q_{0}}f^{(0)}_q\right\} \nonumber \\
     & = & 2\pi\delta(q^{2}-m^{2})(\partial_{\nu}\alpha)\frac{\epsilon^{\mu\nu\beta\alpha}q_{\beta}u_{\alpha}}{2|q_{0}|}f^{(0)}_q(1-f^{(0)}_q).
     \label{eq:Chemical_potential_Leq_AWF}
\end{eqnarray}
The above result is consistent with those obtained using metric perturbation in the imaginary-time formalism \citep{Liu:2020dxg,Liu:2021uhn} and aligns with the massless derivation based on chiral kinetic theory \citep{Hidaka:2018ekt,Yi:2021ryh}.

Similarly, the axial-vector Wigner function related to the shear viscous tensor, i.e. the shear induced polarization, can also be computed, 
\begin{eqnarray}
\delta\mathcal{A}_{{\rm LE},\xi}^{<,\mu}(q,X) & = & -q^{\beta}\xi_{\alpha\beta}(X)2\pi\delta(q^{2}-m^{2})\frac{\epsilon^{\mu\nu\rho\sigma}q_{\rho}u_{\sigma}}{2|q_{0}|}f^{(0)}_q (1-f^{(0)}_q), \label{eq:Shear_Leq_AWF}
\end{eqnarray}
where we have used the Belinfante pseudogauge for energy momentum tensor, such that we
only need to focus on the symmetrized gradient $\xi_{\alpha\beta}=\frac{1}{2}\partial_{(\alpha}\beta_{\beta)}$
known as thermal shear tensor.
More details can be found in the Appendix \ref{sec:A_under_T}.

In this work, we restrict our discussion to the case without interaction corrections. In principle, one can use perturbation theory to calculate interaction corrections in the weakly coupled limit, up to $\mathcal{O}(g^{2})$. Unlike the global equilibrium scenario, where the angular momentum vertex is non-local and challenging to handle, all the correlators discussed in this section can be reliably computed using thermal field theory techniques \citep{Kapusta:2006pm,Bellac:2011kqa}.
It is particularly interesting to evaluate the correlators in Eq.~(\ref{eq:LRT_Leq_corrections}) while accounting for scattering contributions, including both elastic and inelastic processes. This can be achieved using diagrammatic techniques commonly employed in the calculation of transport coefficients within real-time formalism \citep{mahan2013many,Gagnon:2006hi,Gagnon:2007qt}. The only difference here is that one of the vertices is of the axial-vector type, $\gamma^{5}\gamma^{\mu}$. This is the relativistic analogue of calculations for anomalous (or spin) Hall conductivity \citep{sinitsyn2007anomalous,RevModPhys.82.1539,sinova2015spin} in a hot QCD plasma. We expect this to yield the so-called side-jump contribution, up to the leading coupling, as predicted from quantum kinetic theory \citep{Lin:2024zik,Fang:2024vds,Wang:2024lis} through solutions of the spin Boltzmann equation \citep{Fang:2022ttm,Wang:2022yli,Lin:2022tma}.


Now, let us comment on Eqs.~(\ref{eq:Chemical_potential_Leq_AWF}, \ref{eq:Shear_Leq_AWF}). In fact, these equations represent all possible first-order non-interacting local equilibrium corrections to the axial-vector Wigner function. We observe that these results are consistent with those reported in Ref.~\citep{Liu:2021uhn}. This property is rather unique for spin transport compared to electric conductivity in particle number transport. We attribute this distinct feature to the additional $\gamma^{5}\gamma^{\mu}$ factor in the axial-vector current, which makes this result applicable only to the phase-space current.
Notably, these off-equilibrium corrections, up to leading order in the coupling, contribute zero to the axial-vector current $J_{5}^{\mu}(x)$ and are consistent with the $\mathcal{T}$-symmetry analysis. This observation hints at the possible non-renormalization of $J_{5}^{\mu} \sim \mathcal{O}(\partial^{1})$ from off-equilibrium sources.

Before concluding this section, we note that Eqs.(\ref{eq:Chemical_potential_Leq_AWF}, \ref{eq:Shear_Leq_AWF}) are derived based on the the density operator shown in Eq.(\ref{eq:Off-GE_DO}), which extends beyond global equilibrium. As discussed below Eq.~(\ref{eq:LRT_Leq_corrections}), this suggests that the polarization induced by the shear tensor and $\partial_\mu \alpha$ could be dissipative. More systematic studies based on the linear response theory within the Zubarev's approaches becomes necessary and will be present in Ref.~\citep{Fang2:2025}.

\section{Summary and discussion}
\label{sec:summary}

In this work, we analyze the entropy production rate associated with shear-induced polarization and the anomalous Hall effect within the framework of quantum (chiral) kinetic theory. Following the strategy used in classical kinetic theory, we introduce entropy flow and the H-theorem into the chiral kinetic theory. To simplify the complexities of the collision term, we adopt the relaxation time approximation.
We have also verified that the newly introduced entropy flow from chiral kinetic theory is consistent with the one derived from relativistic hydrodynamics, and that the H-theorem ensures entropy increases.

Our analysis reveals that the perturbation of the distribution function around equilibrium, related to the shear viscous tensor, leads to an increase in the entropy production rate, similar to the behavior of the conventional shear viscous tensor in classical kinetic theory. However, both shear-induced polarization and the anomalous Hall effect do not contribute to the entropy production rate after integrating over momentum.
As a result, it remains challenging to definitively conclude whether shear-induced polarization and the anomalous Hall effect are non-dissipative based solely on entropy flow and entropy production rate. These findings highlight the limitations of conventional entropy analysis in capturing quantum effects within the framework of quantum kinetic theory.

We have also studied the shear-induced polarization using linear response theory within the framework of Zubarev's approach. 
The estimation of the spin current operator $\hat{J}_5^\mu$ under time-reversal transformation suggests that shear-induced polarization violates time-reversal symmetry and, therefore, should vanish. However, the analysis on the Wigner function in the phase space based on the Zubarev's approaches indicates that shear-induced polarization is permitted in phase space.
We find that the axial-vector Wigner function $\mathcal{A}^{<,\mu}$ and its variation beyond equilibrium, $\delta \mathcal{A}^{<,\mu}_{\textrm{LE}}$, exhibit the same behavior under time-reversal transformation, as shown in Eqs.~(\ref{eq:Time_reversal_Axial_WF}, \ref{eq:delta_A_under_T}).
Our results indicate that time-reversal symmetry does not impose any constraints on the transport coefficients of the phase-space current associated with shear-induced polarization. Additionally, we have derived $\delta \mathcal{A}^{<,\mu}_{\textrm{LE}}$ induced by the thermal shear tensor and $\partial_\mu \alpha$. According to our derivation, this suggests that the polarization induced by the shear tensor and $\partial_\mu \alpha$ could be dissipative.

Finally, we would like to discuss entropy analysis in quantum kinetic theory. As mentioned, the current studies highlight the limitations of conventional entropy analysis in addressing certain quantum effects within kinetic theory. 
A reasonable extension of the current entropy analysis involves decomposing the distribution function as $f_p = f_{\textrm{GE}} + \delta f_p$, where $f_{\textrm{GE}}$ represents the global equilibrium distribution function, as defined in Sec.~\ref{subsec:Time-reversal-symmetry-analysis} and it can receive interaction corrections even in global equilibrium \citep{Fang:2025pzy}. Roughly speaking, the effects arising from $f_{\textrm{GE}}$ can be considered non-dissipative, while those associated with $\delta f_p$ could be dissipative. This decomposition aligns with the conclusions drawn from Zubarev's approach.
Therefore, to further improve entropy analysis, more systematic studies based on Zubarev's approach are essential. However, such investigations are beyond the scope of this work and will be presented in a forthcoming study.

\begin{acknowledgments}
We are thankful to Francesco Becattini and Matteo Buzzegoli for helpful
discussions. This work is supported in part by the National Key Research
and Development Program of China under Contract No. 2022YFA1605500,
by the Chinese Academy of Sciences (CAS) under Grant No. YSBR-088, by National Nature Science Foundation of China (NSFC) under Grant
No. 12075235 and 12135011, by National Science and Technology Council (Taiwan) under Grant No. NSTC 113-2628-M-001-009-MY4, and by Academia Sinica under Project No. AS-CDA-114-M01.  
\end{acknowledgments}

\appendix

\section{Derivation to Eqs.~(\ref{eq:entropy_without_e}) and (\ref{eq:entropy_production_rate_03})} 
\label{sec:appendix_entropy}

Let us further simplify the last term $-\hbar S_{\left(u\right)}^{\mu\nu}\mathcal{C}_{\nu} \left[\mathcal{H}[f_p^{(n)}]\right]$ in Eq.~(\ref{eq:entropy_without_e}). Since the entropy flow is at $\mathcal{O}(\partial^1)$, we can decompose the collisional corrections $\mathcal{C}_{\nu}$ into two parts, 
\begin{equation}
    \mathcal{C}_{\nu}[\mathcal{H}(f_p^{(n)})] = \mathcal{C}_{\nu}^{leq} + \mathcal{C}_{\nu}^{\delta}, \label{eq:decompose_C_nu}
\end{equation}
where $\mathcal{C}_{\nu}^{leq}\equiv\mathcal{C}_{\nu} \left[\mathcal{H}[f_p^{eq}]\right]$ and $\mathcal{C}_{\nu}^{\delta}\equiv\mathcal{C}_{\nu} \left[\mathcal{H}[\delta f_p]\right]$.
On the other hand, it is naturally to assume that $\mathcal{C}_\nu\left[\mathcal{H}[f]\right] \propto f$. For the first part, $\mathcal{C}_{\nu}^{leq}$ can be parameterize as 
\begin{equation}
    \mathcal{C}_\nu^{leq} =(b_1 u_\nu+b_2 p_\nu + b_3 \partial_\nu^p)f_p^{eq} + b_{4,\nu} f^{(0)}_p,
\end{equation}
where $b_i \sim \mathcal{O}(\partial^0)$, $b_{4,\nu}\sim \mathcal{O}(\partial^1)$ and $f^{(0)}_p$ is given by Eq.~(\ref{eq:distribution_eq}). Here, $\partial_\nu^p\equiv \partial/\partial p^\nu$. 
Inserting the above decomposition into entropy flow, we find the term proportional to $\mathcal{C}_{\nu}^{leq}$ vanishes after the momentum integration. Similarly, we can parameterize $\mathcal{C}_{\nu}^{\delta}$ as 
\begin{equation}
    \mathcal{C}_\nu^\delta = \frac{1}{\tau_R^{\prime\prime}}(a_1 u_\nu+a_2 p_\nu + a_3 \partial_\nu^p)\delta f_p, \label{eq:C_nu_delta}
\end{equation}
where $a_i \sim \mathcal{O}(\partial^0)$ and $\tau_R^{\prime\prime}\sim\mathcal{O}(\partial^0)$ is another relaxation time. In general, $\tau_R^{\prime\prime}$ does not necessary to be the same as $\tau_R$ introduced in Eq.~(\ref{eq:RTA_01}). Inserting the $ \mathcal{C}_\nu^\delta$ into Eq.~(\ref{eq:entropy_without_e}), we can further simplify the entropy flow as
\begin{equation}
    s^{\mu}=2\int\frac{d^{4}p}{\left(2\pi\right)^{3}}\bar{\epsilon}_{(u)}\delta\left(p^{2}\right)\left\{\left(p^{\mu}+\hbar S_{\left(u\right)}^{\mu\nu}\partial_{\nu}\right)\mathcal{H}[f_{p}^{\left(n\right)}]-\hbar \frac{1}{\tau_R^{\prime\prime}} a_3 S_{\left(u\right)}^{\mu\nu} f_p^{eq}(1-f_p^{eq})\partial_\nu^p \chi_p
    \right\}.\label{eq:entropy_without_e_2}
\end{equation}
Integrating by parts, the term proportional to $\partial_\nu^p \chi_p$ in the entropy flow becomes, 
\begin{equation}
    -\int\frac{d^{4}p}{\left(2\pi\right)^{3}}\partial_{\nu}^{p}\left[\bar{\epsilon}_{(u)}\delta\left(p^{2}\right)\hbar\frac{1}{\tau_R^{\prime\prime}}a_{3}f_{p}^{\left(0\right)}\left(1-f_{p}^{\left(0\right)}\right)\right]S_{\left(u\right)}^{\mu\nu}\chi_{p}, \nonumber
\end{equation}
where we have dropped the total derivatives and $\partial_{\nu}^{p}S_{\left(u\right)}^{\mu\nu}=0$. Since both $a_3$ and $\tau_R^\prime$ are scalar and at $\mathcal{O}(\partial^0)$, they can only be the function of $u\cdot p$ and $p_\mu p_\nu \Theta^{\mu\nu}$. After taking the momentum derivative an contracting with $S^{\mu\nu}_{(u)}$, the above integration vanishes. Therefore, the entrop flow reduces to,
\begin{equation}
    s^{\mu}=2\int\frac{d^{4}p}{\left(2\pi\right)^{3}}\bar{\epsilon}_{(u)}\delta\left(p^{2}\right)\left(p^{\mu}+\hbar S_{\left(u\right)}^{\mu\nu}\partial_{\nu}\right)\mathcal{H}[f_{p}^{\left(u\right)}].\label{eq:entropy_without_e_02}
\end{equation}

Next, let us compute the entropy production rate. 
In the chiral kinetic theory with collisions, the current and the energy momentum
tensor in the absence of the external field are given by
\begin{eqnarray}
    J^{\mu}&=&2\int\frac{d^{4}p}{\left(2\pi\right)^{3}}\bar{\epsilon}_{(u)}\delta\left(p^{2}\right)\left\{ \left(p^{\mu}+\hbar S_{\left(u\right)}^{\mu\nu}\partial_{\nu}\right)f_{p}^{\left(u\right)}-\hbar S_{\left(u\right)}^{\mu\nu}\mathcal{C}_{\nu}\left[f_{p}^{(u)}\right]\right\}, \nonumber \\
    T^{\mu\nu} & = & \int\frac{d^{4}p}{\left(2\pi\right)^{3}}\bar{\epsilon}_{(u)}\delta\left(p^{2}\right)\left[p^{\nu}\left\{ \left(p^{\mu}+\hbar S_{\left(u\right)}^{\mu\alpha}\partial_{\alpha}\right)f_{p}^{\left(u\right)}-\hbar S_{\left(u\right)}^{\mu\alpha}\mathcal{C}_{\alpha}\left[f_{p}^{(u)}\right]\right\} \right.\nonumber \\
 &  & \left.+p^{\mu}\left\{ \left(p^{\nu}+\hbar S_{\left(u\right)}^{\nu\alpha}\partial_{\alpha}\right)f_{p}^{\left(u\right)}-\hbar S_{\left(u\right)}^{\nu\alpha}\mathcal{C}_{\alpha}\left[f_{p}^{(u)}\right]\right\} \right],
 \label{eq:J_EMT_CKT_collision}
\end{eqnarray}
Analogous to Eq.~(\ref{eq:decompose_C_nu}), we rewrite the collision term as two parts, $\mathcal{C}_{\alpha}\left[f_{p}^{(n)}\right]=\mathcal{C}_\alpha[f_p^{eq}]+\mathcal{C}_{\alpha}\left[\delta f_{p}\right]$. As mentioned, $\mathcal{C}_{\alpha}\left[f_{p}^{eq}\right]=0$ ensures that Eq.~(\ref{eq:partial_s_classical}) is satisfied. For the part $\mathcal{C}_{\alpha}\left[\delta f_{p}\right]$, we can further parameterize it as 
$\mathcal{C}_{\alpha}\left[\delta f_{p}\right]=\left(b_{1}^\prime u_{\alpha}+b_{2}^\prime p_{\alpha}+b_{3}^\prime \partial_{\alpha}^{p}\right)\delta f_{p}$, where $b_i^\prime \sim \mathcal{O}(\partial^0)$. Inserting this decomposition back to Eqs.~(\ref{eq:J_EMT_CKT_collision}), we find that all terms proportional to $S^{\mu\nu}_{(u)}\mathcal{C}_\nu$ will not contribute.

By using the constrains given by the conservation equations shown in Eqs.~(\ref{eq:constrains_hydro}), we can compute the entropy production rate, 
\begin{eqnarray}
\partial_{\mu}s^{\mu} 
 & = & 2\int\frac{d^{4}p}{\left(2\pi\right)^{3}}\bar{\epsilon}_{(u)}\delta\left(p^{2}\right)\ln\left(\frac{1-f_{p}^{\left(u\right)}}{f_{p}^{\left(u\right)}}\right)\overline{\mathcal{C}}\left[f_{p}^{\left(u\right)}\right]\nonumber \\
 & = & 2\int\frac{d^{4}p}{\left(2\pi\right)^{3}}\bar{\epsilon}_{(u)}\delta\left(p^{2}\right)\left[\beta\left(u\cdot p\right)-\beta\mu+\beta\frac{\hbar p\cdot\omega}{2p\cdot u}-\chi_{p}\right]\overline{\mathcal{C}}\left[f_{p}^{\left(u\right)}\right]\nonumber \\
 & = & 2\beta\int\frac{d^{4}p}{\left(2\pi\right)^{3}}\bar{\epsilon}_{(u)}\delta\left(p^{2}\right)\left\{ \frac{\hbar}{2}\left[\left(p\cdot u\right)\left(\partial_{\mu}S_{\left(u\right)}^{\mu\alpha}\right)\partial_{\alpha}+p^{\mu}\left(\partial_{\mu}u_{\nu}\right)S_{\left(u\right)}^{\nu\alpha}\partial_{\alpha}\right]f_{p}^{\left(u\right)}\right.\nonumber \\
 &  & \left.+\frac{\hbar p\cdot\omega}{2p\cdot u}\overline{\mathcal{C}}\left[f_{p}^{\left(u\right)}\right]-\frac{1}{\beta}\chi_{p}\overline{\mathcal{C}}\left[f_{p}^{\left(u\right)}\right]\right\} \nonumber \\
 & \simeq & -2\int\frac{d^{4}p}{\left(2\pi\right)^{3}}\bar{\epsilon}_{(u)}\delta\left(p^{2}\right)\chi_{p}\overline{\mathcal{C}}\left[f_{p}^{\left(u\right)}\right] + \mathcal{O}(\partial^3),
\end{eqnarray}
where we have used the relation 
$\left(p\cdot u\right)\left(\partial_{\mu}S_{\left(u\right)}^{\mu\alpha}\right)+p^{\mu}\left(\partial_{\mu}u_{\nu}\right)S_{\left(u\right)}^{\nu\alpha}=-\frac{p\cdot\omega}{p\cdot u}p^{\alpha}$ and have used the spin Boltzmann equation (\ref{eq:CKT_RTA}) in the last step. 
Inserting the collision term in Eq.~(\ref{eq:RTA_01}) into above equations, yields, 
\begin{equation}
    \partial_\mu s^\mu = I_1 + I_2,
\end{equation}
where
\begin{eqnarray}
I_1 & = & \frac{2}{\tau_{R}}\int\frac{d^{4}p}{\left(2\pi\right)^{3}}\bar{\epsilon}_{(u)}\delta\left(p^{2}\right)\chi_{p}\left(p\cdot u\right)\delta f_{p}\nonumber \\
& = & \frac{1}{\tau_{R}}\int\frac{d^{3}p}{\left(2\pi\right)^{3}}\left\{ \left.\left[f_{p}^{eq}\left(1-f_{p}^{eq}\right)\chi_{p}^{2}\right]\right|_{p_{0}=\left|\mathbf{p}\right|}+\left.\left[f_{p}^{eq}\left(1-f_{p}^{eq}\right)\chi_{p}^{2}\right]\right|_{p_{0}=-\left|\mathbf{p}\right|}\right\} \geq0.\nonumber \\
\end{eqnarray}
and
\begin{eqnarray}
I_2 & = & \frac{2\hbar}{\tau_{R}}\int\frac{d^{4}p}{\left(2\pi\right)^{3}}\bar{\epsilon}_{(u)}\delta\left(p^{2}\right)\chi_{p}\frac{p^{\mu}\mathcal{R}_{\mu}}{\left(p\cdot u\right)^{2}}\delta f_{p}.\label{eq:entropy_prod_I2}
\end{eqnarray}
Inserting the parameterization for $\mathcal{R}^\mu$ in Eq.~(\ref{eq:R_parameterization}), yields the Eq.~(\ref{eq:entropy_production_rate_03}).

\section{Derivation of Eqs.~(\ref{eq:Time_reversal_Axial_WF}, \ref{eq:delta_A_under_T}, \ref{eq:Shear_Leq_AWF})}
\label{sec:A_under_T}

We first consider the $\mathcal{A}^{<,\mu}(q,X)$ under time reversal transformation,
\begin{eqnarray}
 &  & \hat{T}^{-1}\mathcal{A}^{<,\mu}(q,X)\hat{T}\nonumber \\
 & = & -\frac{1}{4}\left(\hat{T}^{-1}\int{\rm d}^{4}Y\hat{T}\right)\left(\hat{T}^{-1}e^{iq\cdot Y}\hat{T}\right)\left(\hat{T}^{-1}{\rm Tr}\left(\hat{\rho}\overline{\psi}(y)\gamma^{5}\gamma_{\mu}\psi(x)\right)\hat{T}\right)\nonumber \\
 & = & -\frac{1}{4}\left((-1)\int{\rm d}^{4}Y\right)e^{iq\cdot Y}\left[(-1)^{\mu}\left(\overline{\psi}(-y_{0},\boldsymbol{y})\gamma^{5}\gamma_{\mu}\psi(-x_{0},\boldsymbol{x})\right)\right]\nonumber \\
 & = & -\frac{1}{4}(-1)(-1)^{\mu}\int_{-\infty}^{+\infty}{\rm d}(x_{0}-y_{0})\int{\rm d}^{3}(\boldsymbol{x}-\boldsymbol{y})e^{-iq_{0}\left(-x_{0}-(-y_{0})\right)-i\boldsymbol{q}\cdot(\boldsymbol{x}-\boldsymbol{y})}\overline{\psi}(-y_{0},\boldsymbol{y})\gamma^{5}\gamma_{\mu}\psi(-x_{0},\boldsymbol{x})\nonumber \\
 & = & -\frac{1}{4}(-1)(-1)^{\mu}\int_{-\infty}^{+\infty}{\rm d}(x_{0}^{\prime}-y_{0}^{\prime})\int{\rm d}^{3}(\boldsymbol{x}-\boldsymbol{y})e^{-iq_{0}(x_{0}^{\prime}-y_{0}^{\prime})-i\boldsymbol{q}\cdot(\boldsymbol{x}-\boldsymbol{y})}\overline{\psi}(y_{0}^{\prime},\boldsymbol{y})\gamma^{5}\gamma_{\mu}\psi(x_{0}^{\prime},\boldsymbol{x})\nonumber \\
 & = & (-1)(-1)^{\mu}\mathcal{A}^{<,\mu}(-q_{0},\boldsymbol{q};\frac{x_{0}^{\prime}+y_{0}^{\prime}}{2},\boldsymbol{X})\nonumber \\
 & = & -(-1)^{\mu}\mathcal{A}^{<,\mu}(-q_{0},\boldsymbol{q};-X_{0},\boldsymbol{X}),
 \label{eq:Time_reversal_Axial_WF_detials}
\end{eqnarray}
where $q$ transforms as a 4-momentum and $Y$ a position vector. Here,
we also introduce the notation $x_{0}^{\prime}=-x_{0}$ and $y_{0}^{\prime}=-y_{0}$
in the forth step. Note that the $\mathcal{T}$-transformation of $J_{5}^{\mu}(X)$ can
be recovered, 
\begin{eqnarray}
\hat{T}^{-1}J_{5}^{\mu}(X)\hat{T} & = & \hat{T}^{-1}\left(4\int\frac{{\rm d}^{4}q}{(2\pi)^{4}}\mathcal{A}^{<,\mu}(q,X)\right)\hat{T}=4(-1)^{3}\int\frac{{\rm d}^{4}q}{(2\pi)^{4}}\left[-(-1)^{\mu}\mathcal{A}^{<,\mu}(-q_{0},\boldsymbol{q};-X_{0},\boldsymbol{X})\right]\nonumber \\
 & = & (-1)^{\mu}4\int\frac{{\rm d}q_{0}{\rm d}^{3}\boldsymbol{q}}{(2\pi)^{4}}\mathcal{A}^{<,\mu}(q_{0},\boldsymbol{q};-X_{0},\boldsymbol{X})=(-1)^{\mu}J_{5}^{\mu}(-X_{0},\boldsymbol{X}). \label{eq:Time_reversal_J5_details}
\end{eqnarray}

Next, under time-reversal transformation
$\mathcal{T}$, we have,
\begin{eqnarray}
 &  & \mathcal{T}\delta\mathcal{A}_{{\rm LE},\alpha}^{<,\mu}(q,X)\nonumber \\
 & = & \left(\hat{T}^{-1}\partial_{\nu}\alpha(X)\hat{T}\right)\lim_{k\to0}{\rm Im}\left(\hat{T}^{-1}\frac{1}{\beta(X)k_{0}}\frac{1}{4}\hat{T}\right)\left(T^{-1}\int{\rm d}^{4}Ye^{iq\cdot Y}\int_{-\infty}^{+\infty}{\rm d}^{4}x^{\prime}e^{-ik\cdot(x^{\prime}-X)}T\right)\nonumber \\
 &  & \;\times\left(\hat{T}^{-1}(-i)\theta(X_{0}-x_{0}^{\prime})\hat{T}\right)\left(\hat{T}^{-1}\left\langle \hat{\overline{\psi}}(y)\gamma^{5}\gamma^{\mu}\hat{\psi}(x)\hat{j}^{\nu}(x^{\prime})-\hat{j}^{\nu}(x^{\prime})\hat{\overline{\psi}}(y)\gamma^{5}\gamma^{\mu}\hat{\psi}(x)\right\rangle _{{\rm GE}}\hat{T}\right)\nonumber \\
 & = & (-1)^{2}(-1)^{\nu}(-1)\partial_{\nu}\alpha(-X_{0},\boldsymbol{X})\lim_{k\to0}{\rm Im}\frac{1}{\beta(-X_{0},\boldsymbol{X})k_{0}}\frac{1}{4}\int{\rm d}^{4}Ye^{iq\cdot Y}\nonumber \\
 &  & \;\times\int_{-\infty}^{+\infty}{\rm d}^{4}x^{\prime}e^{-ik\cdot(x^{\prime}-X)}(-1)(-i)\theta(-X_{0}+x_{0}^{\prime})(-1)^{\mu+\nu}\nonumber \\
 &  & \;\times\left\langle \hat{\overline{\psi}}(-y_{0},\boldsymbol{y})\gamma^{5}\gamma^{\mu}\hat{\psi}(-x_{0},\boldsymbol{x})\hat{j}^{\nu}(-x_{0}^{\prime},\boldsymbol{x}^{\prime})-\hat{j}^{\nu}(-x_{0}^{\prime},\boldsymbol{x}^{\prime})\hat{\overline{\psi}}(-y_{0},\boldsymbol{y})\gamma^{5}\gamma^{\mu}\hat{\psi}(-x_{0},\boldsymbol{x})\right\rangle _{{\rm GE}}\nonumber \\
 & = & (-1)(-1)^{\mu}\partial_{\nu}\alpha(-X_{0},\boldsymbol{X})\lim_{k_{0}\to0}\lim_{k_{i}\to0}{\rm Im}\frac{1}{4\beta(-X_{0},\boldsymbol{X})}\int{\rm d}^{4}Ye^{iq\cdot Y}\nonumber \\
 &  & \;\times\int_{-\infty}^{+\infty}{\rm d}^{4}x^{\prime}(x_{0}^{\prime}-X_{0})e^{-ik\cdot(x^{\prime}-X)}(-i)\theta(-X_{0}+x_{0}^{\prime})\nonumber \\
 &  & \;\times\left\langle \hat{\overline{\psi}}(-y_{0},\boldsymbol{y})\gamma^{5}\gamma^{\mu}\hat{\psi}(-x_{0},\boldsymbol{x})\hat{j}^{\nu}(-x_{0}^{\prime},\boldsymbol{x}^{\prime})-\hat{j}^{\nu}(-x_{0}^{\prime},\boldsymbol{x}^{\prime})\hat{\overline{\psi}}(-y_{0},\boldsymbol{y})\gamma^{5}\gamma^{\mu}\hat{\psi}(-x_{0},\boldsymbol{x})\right\rangle _{{\rm GE}},
\end{eqnarray}
where 
\begin{eqnarray}
    \hat{T}^{-1}j_{5}^{\mu}(x)j^{\nu}(y)\hat{T} & = & (-1)^{\mu+\nu}j_{5}^{\mu}(-x_{0},\boldsymbol{x})j^{\nu}(-y_{0},\boldsymbol{y}).
\end{eqnarray}
We may alternatively start from
\begin{eqnarray}
 &  & \delta\mathcal{A}_{{\rm LE},\alpha}^{<,\mu}(-q_{0},\boldsymbol{q};-X_{0},\boldsymbol{X})\nonumber \\
 & = & \partial_{\nu}\alpha(-X_{0},\boldsymbol{X})\lim_{k_{0}\to0}\lim_{k_{i}\to0}{\rm Im}\frac{1}{\beta(-X_{0},\boldsymbol{X})k_{0}}\frac{1}{4}\int{\rm d}^{4}Ye^{-iq_{0}Y_{0}}e^{-i\boldsymbol{q}\cdot\boldsymbol{Y}}\nonumber \\
 &  & \;\times\int_{-\infty}^{+\infty}{\rm d}^{4}x^{\prime}e^{-ik_{0}(x_{0}^{\prime}+X_{0})}e^{i\boldsymbol{k}\cdot(\boldsymbol{x}^{\prime}-\boldsymbol{X})}(-i)\theta(-X_{0}-x_{0}^{\prime})\nonumber \\
 &  & \;\times\left\langle \hat{\overline{\psi}}_{b}(-X_{0}-\frac{Y_{0}}{2},\boldsymbol{y})\gamma^{5}\gamma^{\mu}\hat{\psi}(-X_{0}+\frac{Y_{0}}{2},\boldsymbol{x})\hat{j}^{\nu}(x_{0}^{\prime},\boldsymbol{x}^{\prime})\right.\nonumber \\
 &  & \;\left.-\hat{j}^{\nu}(x_{0}^{\prime},\boldsymbol{x}^{\prime})\hat{\overline{\psi}}(-X_{0}-\frac{Y_{0}}{2},\boldsymbol{y})\gamma^{5}\gamma^{\mu}\hat{\psi}(-X_{0}+\frac{Y_{0}}{2},\boldsymbol{x})\right\rangle _{{\rm GE}}\nonumber \\
 & = & \partial_{\nu}\alpha(-X_{0},\boldsymbol{X})\lim_{k_{i}\to0}{\rm Im}\frac{1}{4\beta(-X_{0},\boldsymbol{X})}\int{\rm d}^{4}Ye^{iq\cdot Y}\nonumber \\
 &  & \;\times\int_{-\infty}^{+\infty}{\rm d}^{4}x^{\prime}i(x_{0}^{\prime}-X_{0})\lim_{k_{0}\to0}e^{ik_{0}(x_{0}^{\prime}-X_{0})}e^{i\boldsymbol{k}\cdot(\boldsymbol{x}^{\prime}-\boldsymbol{X})}(-i)\theta(-X_{0}+x_{0}^{\prime})\nonumber \\
 &  & \;\times\left\langle \hat{\overline{\psi}}_{b}(-y_{0},\boldsymbol{y})\gamma^{5}\gamma^{\mu}\hat{\psi}(-x_{0},\boldsymbol{x})\hat{j}^{\nu}(-x_{0}^{\prime},\boldsymbol{x}^{\prime})-\hat{j}^{\nu}(-x_{0}^{\prime},\boldsymbol{x}^{\prime})\hat{\overline{\psi}}(-y_{0},\boldsymbol{y})\gamma^{5}\gamma^{\mu}\hat{\psi}(-x_{0},\boldsymbol{x})\right\rangle _{{\rm GE}}.
\end{eqnarray}
Namely, we recover the behavior of axial-vector current under time-reversal
transformation derived in Eq.~(\ref{eq:Time_reversal_Axial_WF}).

Next, we will compute the shear induced polarization for Eq.~(\ref{eq:CPG_induced_polar}) 
Substituting the integral representation of
sign function,
\begin{eqnarray}
\theta(X_{0}-x_{0}^{\prime}) & = & i\int_{-\infty}^{+\infty}\frac{\mathrm{d}k_{0}^{\prime}}{2\pi}\frac{e^{-ik_{0}^{\prime}(X_{0}-x_{0}^{\prime})}}{k_{0}^{\prime}+i\eta},
\end{eqnarray}
Eq.~(\ref{eq:CPG_induced_polar}) reduces to,
\begin{eqnarray}
 &  & \delta\mathcal{A}_{{\rm LE},\alpha}^{<,\mu}(q,X)\nonumber \\
 & = & \partial_{\nu}\alpha(X)\lim_{k\to0}{\rm Im}\frac{T(X)}{k_{0}}\frac{1}{4}\int{\rm d}^{4}Ye^{iq\cdot Y}\int_{-\infty}^{+\infty}{\rm d}^{4}x^{\prime}\int_{-\infty}^{+\infty}\frac{\mathrm{d}k_{0}^{\prime}}{2\pi}\frac{e^{-ik_{0}^{\prime}(X_{0}-x_{0}^{\prime})}}{k_{0}^{\prime}+i\eta}e^{-ik\cdot(x^{\prime}-X)}\nonumber \\
 &  & \;\times{\rm Tr}\left\{ \gamma^{5}\gamma^{\mu}S_{0,{\rm GE}}^{>}(X+\frac{Y}{2},x^{\prime})\gamma^{\nu}S_{0,{\rm GE}}^{<}(x^{\prime},X-\frac{Y}{2})-\gamma^{5}\gamma^{\mu}S_{{\rm 0,GE}}^{<}(X+\frac{Y}{2},x^{\prime})\gamma^{\nu}S_{0,{\rm GE}}^{>}(x^{\prime},X-\frac{Y}{2})\right\} .\nonumber \\
\label{eq:CPG_induced_polar_1}
\end{eqnarray}
We emphasize the $\theta$-function here only accounts for the boundary
condition and the small imaginary part in the denominator should be
taken to zero in the last. Here we have also introduced the following
global equilibrium Wigner functions for the free fermions as follows
\begin{eqnarray}
S_{0,{\rm GE}}^{\lessgtr}(x,y) & = & \int\frac{{\rm d}^{4}q}{(2\pi)^{4}}e^{-iq\cdot(x-y)}S_{0,{\rm GE}}^{\lessgtr}(q,\frac{x+y}{2}),\\
S_{0,{\rm GE}}^{\lessgtr}(q,X) & = & 2\pi\overline{\epsilon}_{(u)}\delta(q^{2}-m^{2})(\gamma^{\mu}q_{\mu}+m)f^{\lessgtr,(0)}(q),\label{eq:Geq_FWF}
\end{eqnarray}
where up to $\mathcal{O}(\partial^{0})$, the free fermion distribution
function $f^{(0)}(q)$ is same for massless and massive fermions and
defined in Eq.(\ref{eq:distribution_eq}) and $\overline{\epsilon}_{(u)}$ is the
sign function. 
For convenience, we also introduce $f^> = 1 -f^<$.
We will work in the $\beta$-frame \citep{Becattini:2014yxa}
with fluid velocity $u^{\mu}=\beta^{\mu}/\sqrt{\beta^{2}}$. Then
Eq.(\ref{eq:CPG_induced_polar_1}) reduces to 
\begin{eqnarray}
 &  & \delta\mathcal{A}_{{\rm LE},\alpha}^{<,\mu}(q,X)\nonumber \\
 & = & \partial_{\nu}\alpha(X)\lim_{k\to0}{\rm Im}\frac{T(X)}{k_{0}}\frac{1}{4}\int\frac{{\rm d}^{4}p_{1}}{(2\pi)^{4}}\frac{{\rm d}^{4}p_{2}}{(2\pi)^{4}}\int_{-\infty}^{+\infty}\frac{\mathrm{d}k_{0}^{\prime}}{2\pi}\frac{1}{k_{0}^{\prime}+i\eta}2\pi\delta(k_{0}+p_{2,0}-p_{1,0}-k_{0}^{\prime})\nonumber \\
 &  & \;\times(2\pi)^{3}\delta^{(3)}(\boldsymbol{k}+\boldsymbol{p}_{2}-\boldsymbol{p}_{1})(2\pi)^{4}\delta^{(4)}(q-\frac{p_{1}+p_{2}}{2})\nonumber \\
 &  & \;\times{\rm Tr}\left\{ \gamma^{5}\gamma^{\mu}S_{0}^{>}(p_{1})\gamma^{\nu}S_{0}^{<}(p_{2})-\gamma^{5}\gamma^{\mu}S_{0}^{<}(p_{1})\gamma^{\nu}S_{0}^{>}(p_{2})\right\} \nonumber \\
 & = & \partial_{\nu}\alpha(X)\lim_{k\to0}{\rm Im}\frac{T(X)}{k_{0}}\frac{1}{4}\int\frac{{\rm d}^{4}Q}{(2\pi)^{4}}\frac{1}{k_{0}-Q_{0}+i\eta}(2\pi)^{3}\delta^{(3)}(\boldsymbol{k}-\boldsymbol{Q})\nonumber \\
 &  & \;\times{\rm Tr}\left\{ \gamma^{5}\gamma^{\mu}S_{0}^{>}(q+\frac{Q}{2})\gamma^{\nu}S_{0}^{<}(q-\frac{Q}{2})-\gamma^{5}\gamma^{\mu}S_{0}^{<}(q+\frac{Q}{2})\gamma^{\nu}S_{0}^{>}(q-\frac{Q}{2})\right\}.
\end{eqnarray}
Inserting Eq.(\ref{eq:Geq_FWF}) into above equation, yields
\begin{eqnarray}
 &  & \delta\mathcal{A}_{{\rm LE},\alpha}^{<,\mu}(q,X)\nonumber \\
 & = & \partial_{\nu}\alpha(X)\lim_{k_{0}\to0}\frac{\partial}{\partial k_{0}}\lim_{k_{i}\to0}{\rm Im}\frac{T(X)}{k_{0}}\frac{1}{4}\int\frac{{\rm d}^{4}Q}{(2\pi)^{2}}\frac{1}{k_{0}-Q_{0}+i\eta}(2\pi)^{3}\delta^{(3)}(\boldsymbol{k}-\boldsymbol{Q})\delta\left((q+\frac{Q}{2})^{2}-m^{2}\right)\nonumber \\
 &  & \;\times\delta\left((q-\frac{Q}{2})^{2}-m^{2}\right)\left\{ f^{<,(0)}(q-\frac{Q}{2})f^{>,(0)}(q+\frac{Q}{2})-f^{<,(0)}(q+\frac{Q}{2})f^{>,(0)}(q-\frac{Q}{2})\right\} \nonumber \\
 &  & \;\times{\rm Tr}\left[\gamma^{5}\gamma^{\mu}(\gamma^{\alpha}(q+\frac{Q}{2})_{\alpha}+m)\gamma^{\nu}(\gamma^{\beta}(q-\frac{Q}{2})_{\beta}+m)\right]\nonumber \\
 & = & \partial_{\nu}\alpha(X)\lim_{k_{0}\to0}\lim_{k_{i}\to0}\frac{T(X)}{k_{0}}{\rm Im}\frac{1}{4}\int\frac{{\rm d}^{4}Q}{(2\pi)^{2}}\frac{1}{k_{0}-Q_{0}+i\eta}(2\pi)^{3}\delta^{(3)}(\boldsymbol{k}-\boldsymbol{Q})(-4i\epsilon^{\mu\alpha\nu\beta}Q_{\alpha}q_{\beta})\nonumber \\
 &  & \;\times\delta(q^{2}+\frac{Q^{2}}{4}-m^{2})\delta(-2q\cdot Q)\left\{ f^{<,(0)}(q-\frac{Q}{2})-f^{<,(0)}(q+\frac{Q}{2})\right\} \nonumber \\
 & = & \partial_{\nu}\alpha(X)\lim_{k_{0}\to0}\frac{T(X)}{k_{0}}\lim_{k_{i}\to0}{\rm Im}\int\frac{{\rm d}Q_{0}}{2\pi}(2\pi)^{2}\frac{\delta(Q_{0})Q_{0}^{2}}{k_{0}-Q_{0}+i\eta}(-i\epsilon^{\mu\alpha\nu\beta}q_{\beta}u_{\alpha})\nonumber \\
 &  & \;\times\delta(q^{2}+\frac{Q_{0}^{2}}{4}-m^{2})\frac{1}{2|q_{0}|}\left\{ -\partial_{q_{0}}f^{<,(0)}(q)+\mathcal{O}(Q_{0}^{2})\right\} ,
\end{eqnarray}
where $q_{0}=u\cdot q$. In the $k\to0$ limit, 
\begin{eqnarray}
\delta\mathcal{A}_{{\rm LE},\alpha}^{<,\mu}(q,X) & = & T(X)\partial_{\nu}\alpha(X)\int\frac{{\rm d}Q_{0}}{(2\pi)}\delta(Q_{0})Q_{0}^{2}\frac{Q_{0}^{2}-\eta^{2}}{(Q_{0}^{2}+\eta^{2})^{2}}\epsilon^{\mu\alpha\nu\beta}q_{\beta}u_{\alpha}\nonumber \\
 &  & \;\times(2\pi)^{2}\delta(q^{2}+\frac{Q_{0}^{2}}{4}-m^{2})\frac{1}{2|q_{0}|}\left\{ -\partial_{q_{0}}f^{<,(0)}(q)+\mathcal{O}(Q_{0}^{2})\right\} ,
\end{eqnarray}
taking the $\eta\to0^{+}$ introduced as boundary constraint for the
$\theta$-function, we obtain,
\begin{equation}
    \delta\mathcal{A}_{{\rm LE},\alpha}^{<,\mu}(q,X)  =  2\pi\delta(q^{2}-m^{2})T(\partial_{\nu}\alpha)\epsilon^{\mu\alpha\nu\beta}q_{\beta}u_{\alpha}\frac{1}{2|q_{0}|}\left\{ -\partial_{q_{0}}f^{(0)}_q\right\},
\end{equation}
which reduces to Eq.~(\ref{eq:Chemical_potential_Leq_AWF}).
It is found that the infinitesimal $\eta$ only plays a role of imposing
proper initial conditions for the retarded correlators, it can be simply taken zero in the calculation of the correlators.
Such observation is also noticed when calculating the transport coefficients
e.g. in Ref.~\citep{Jeon:1994if}.

The axial-vector Wigner function related to the shear viscous tensor is given by in the Belinfante pseudo gauge, 
\begin{eqnarray}
\delta\mathcal{A}_{{\rm LE},\xi}^{<,\mu}(q,X) & = & -\frac{1}{8}\xi_{\alpha\beta}(X)\lim_{k_{0}\to0}\lim_{k_{i}\to0}\frac{T(X)}{k_{0}}{\rm Im}\int_{-\infty}^{+\infty}{\rm d}^{4}x^{\prime}i\int_{-\infty}^{+\infty}\frac{{\rm d}k_{0}^{\prime}}{2\pi}\frac{e^{-ik_{0}^{\prime}(X_{0}-t^{\prime})}}{k_{0}^{\prime}+i\eta}e^{-ik\cdot(x^{\prime}-X)}\nonumber \\
 &  & \;\times\int{\rm d}^{4}ye^{iq\cdot y}\left\{ (-1)^{2}{\rm Tr}\left[\gamma^{5}\gamma^{\mu}S^{>}(X+\frac{y}{2},x^{\prime})\gamma^{\alpha}(\overrightarrow{\partial}_{x^{\prime}}^{\beta}-\overleftarrow{\partial}_{x^{\prime}}^{\beta})S^{<}(x^{\prime},X-\frac{y}{2})\right]\right.\nonumber \\
 &  & \;\;\left.-(-1)^{2}{\rm Tr}\left[\gamma^{5}\gamma^{\mu}S^{<}(X+\frac{y}{2},x^{\prime})\gamma^{\alpha}(\overrightarrow{\partial}_{x^{\prime}}^{\beta}-\overleftarrow{\partial}_{x^{\prime}}^{\beta})S^{>}(x^{\prime},X-\frac{y}{2})\right]\right\}.
\end{eqnarray}
We now insert the Fourier components,
\begin{eqnarray}
\delta\mathcal{A}_{{\rm LE},\xi}^{<,\mu}(q,X) & = & \frac{1}{4}q^{\beta}\xi_{\alpha\beta}(X)\lim_{k_{0}\to0}\lim_{k_{i}\to0}\frac{T(X)}{k_{0}}{\rm Im}\int\frac{{\rm d}^{4}Q}{(2\pi)^{4}}\frac{1}{k_{0}-Q_{0}+i\eta}(2\pi)^{3}\delta^{(3)}(\boldsymbol{k}-\boldsymbol{Q})\nonumber \\
 &  & \;\times\left\{ -{\rm Tr}\left[\gamma^{5}\gamma^{\mu}S^{>}(q+\frac{Q}{2})\gamma^{\alpha}S^{<}(q-\frac{Q}{2})\right]+{\rm Tr}\left[\gamma^{5}\gamma^{\mu}S^{<}(q+\frac{Q}{2})\gamma^{\alpha}S^{>}(q-\frac{Q}{2})\right]\right\}.\nonumber \\
\end{eqnarray}
It is interesting to notice the kernel is same as the $\mathcal{A}_{\alpha}$
sector, where
\begin{eqnarray}
 &  & \lim_{k_{0}\to0}\lim_{k_{i}\to0}{\rm Im}\frac{T(X)}{k_{0}}\frac{1}{4}\int\frac{{\rm d}^{4}Q}{(2\pi)^{4}}\frac{1}{k_{0}-Q_{0}+i\eta}(2\pi)^{3}\delta^{(3)}(\boldsymbol{k}-\boldsymbol{Q})\nonumber \\
 &  & \;\times{\rm Tr}\left\{ \gamma^{5}\gamma^{\mu}S^{>}(q+\frac{Q}{2})\gamma^{\nu}S^{<}(q-\frac{Q}{2})-\gamma^{5}\gamma^{\mu}S^{<}(q+\frac{Q}{2})\gamma^{\nu}S^{>}(q-\frac{Q}{2})\right\} \nonumber \\
 & = & 2\pi\delta(q^{2}-m^{2})\frac{\epsilon^{\mu\nu\beta\alpha}q_{\beta}u_{\alpha}}{2|q_{0}|}f^{<,(0)}(q)f^{>,(0)}(q),
\end{eqnarray}
which gives Eq.~(\ref{eq:Shear_Leq_AWF}).

\bibliographystyle{h-physrev}
\bibliography{qkt-ref20230407}

\end{document}